\documentclass{article}
\usepackage{algorithm}
\usepackage[noend]{algpseudocode}
\usepackage{authblk}
\usepackage{graphicx}
\usepackage{xcolor}
\usepackage{fullpage} 
\usepackage{hyperref}
\usepackage{amsmath} 
\usepackage{amssymb}
\usepackage{amsthm}
\RequirePackage[T1]{fontenc} 
\RequirePackage[tt=false, type1=true]{libertine}
\RequirePackage[varqu]{zi4}
\RequirePackage[libertine]{newtxmath}
\usepackage{algorithm}
\usepackage{algpseudocode}
\hypersetup{
    colorlinks,
    linkcolor={blue},
    citecolor={blue},
    urlcolor={blue}
}

\newtheorem{theorem}{Theorem}
\newtheorem{definition}{Definition}

\newtheorem{lemma}{Lemma}
\newtheorem{corollary}{Corollary}

\title{{\tt pylspack}: Parallel algorithms and data structures for sketching, column subset selection, regression and leverage scores}

\author[1]{Aleksandros Sobczyk}
\affil[1]{{\small obc@zurich.ibm.com, IBM Research Europe and ETH Zürich, Zürich, Switzerland}}
\author[2]{Efstratios Gallopoulos}
\affil[2]{{\small stratis@ceid.upatras.gr, HPCLAB, Computer Engineering and Informatics Dept., University of Patras, Patras, Greece}}

\date{}

\begin{document}

\maketitle

\begin{abstract}  

{We present parallel algorithms and data structures for three fundamental operations in Numerical Linear Algebra: (i) Gaussian and CountSketch random projections and their combination, (ii) computation of the Gram matrix  and (iii) computation of the squared row norms of the product of two matrices, with a special focus on ``tall-and-skinny'' matrices, which arise in many applications. We provide a detailed analysis of the ubiquitous CountSketch transform and its combination with Gaussian random projections, accounting for memory requirements, computational complexity and workload balancing. {We also demonstrate how these results can be  applied to} column subset selection, least squares regression and leverage scores computation. 
These tools have been implemented in {\tt pylspack}, a publicly available  Python package,\footnote{\hyperlink{https://github.com/IBM/pylspack}{https://github.com/IBM/pylspack}} whose core is written in C++ and parallelized with OpenMP, and which is compatible with standard matrix data structures of SciPy and NumPy. Extensive numerical experiments indicate  that the proposed algorithms scale well and significantly outperform existing libraries for tall-and-skinny matrices.
}
\end{abstract}

\section{Introduction\label{sec:introduction}}

In the past two decades, randomized dimensionality reduction, also known as ``sketching'', {has become a fundamental operation in Randomized Numerical Linear Algebra (RandNLA)}.  The underlying concept is to use random linear maps to reduce the problem size and ultimately derive,  with high probability, fast exact or approximate solutions. This is typically achieved by random sampling or random projections. Such approaches have been studied in the context of applications in several areas, including {Graph Algorithms} \cite{Spielman:2008ka,avron2010counting}, Machine Learning \cite{gittens2013revisiting,cohen2015dimensionality}, Optimization \cite{pilanci2017newton,derezinski2021newton} and Quantum Mechanics \cite{kontopoulou2020randomized}.

A variety of random projections have been proposed in the literature, with different structures, sparsity and statistical guarantees; cf. \cite{achlioptas2001database,indyk1998approximate,ailon2006approximate,ailon2009fast,tropp2011improved,nelson2013sparsity,nelson2013osnap,clarkson2013low_jacm,kane2014sparser,kapralov2016fake,meng2013low,cohen2016nearly,derezinski2021sparse,cartis2021hashing} to name a few. In applications, such transforms often satisfy the Johnson-Lindenstrauss lemma \cite{johnson1984extensions} or provide Oblivious Subspace Embeddings \cite{sarlos2006improved}. {For convenience, we recall these concepts.}
\begin{definition}
	A random matrix $S \in \mathbb{R}^{r\times d}$ is a Johnson-Lindenstrauss transform (JLT) with parameters $\epsilon,\delta,n$, or
	$(\epsilon,\delta)$-JLT for short, if with probability at least $1-\delta$, for any $n$-element set $V \subseteq \mathbb{R}^{d}$ it holds that 
$|(S{ v})^\top Sw-v^\top w |\leq\epsilon \| v\|\| w\|$ for any $ v, w \in V$.
	\label{def:JLT}
\end{definition}
\begin{definition}
	A random matrix $S \in \mathbb{R}^{r\times n}$ is an $(\epsilon,\delta)$-Oblivious Subspace Embedding ($(\epsilon,\delta)$-OSE) with parameters $\epsilon,\delta\in(0,1)$, if with probability at least $1-\delta$, for any fixed matrix $A\in\mathbb{R}^{n\times d}$, $n>d$, it holds that 
$
(1-\epsilon)\| x\|\leq \|S x\| \leq (1+\epsilon)\| x\|,
$ for {all} $ x \in$ range$(A)$ with probability at least $1-\delta$.
\label{def:OSE}
\end{definition}

{From these definitions, it is clear that specialized matrix-multiplication related operations are a key component for sketching to be useful in practice. As noted in \cite{martinsson2020randomized}, such operations must take into account special structures and face the challenges of sparse computations, including efficient data structures for representing the underlying matrices as well as specialized algorithms.  
Such aspects were mentioned, but not elaborated, in our recent companion paper \cite{sobczyk2021estimating}. We here aim to contribute in this direction, by providing a detailed analysis for a set of specific operations and sketching transforms which are critical for various important problems in RandNLA and beyond, for example leverage scores computation \cite{drineas2012fast,clarkson2013low_jacm,nelson2013osnap,alaoui2015fast,Li:2013is,cohen2017input,Cohen:2015cb,zuo2021quantum,sobczyk2021estimating}, least squares regression \cite{sarlos2006improved,nelson2013osnap,woodruff2014sketching,RokhlinTygert.08,Drineas2010,avron2010blendenpik,meng2014lsrn,pilanci2016iterative}, column subset selection \cite{cheung2013fast,boutsidis2009improved, avron2013faster,civril2014column,shitov2021column,sobczyk2021estimating}, and rank computation \cite{saunders2004matrix,cheung2013fast,chepurko2022near,bafna2021optimal,meier2021fast}. 
Specifically, given a tall-and-skinny matrix  $A\in\mathbb{R}^{n\times d}$ with $n\gg d$, we present parallel algorithms, data structures and a numerical library for the following set of operations, all of which can be seen as special cases of sparse matrix multiplication:
\begin{align*}
\begin{array}{c c c c c}
(i)\  SA, &
(ii)\ GA, &
(iii)\ GSA, &
(iv)\ A^\top A, &
(v)\ \text{diag}(ABB^\top A^\top),
\end{array}
\end{align*}
where $G$ is a matrix with i.i.d elements from the standard normal distribution, $B$ is an arbitrary dense matrix and $S$ is the CountSketch matrix \cite{charikar2002finding,clarkson2013low_jacm,nelson2013osnap}.} We we remind its definition from \cite{nelson2013osnap}:
\begin{definition}[\cite{nelson2013osnap}]
A CountSketch matrix $S\in\mathbb{R}^{r\times n}$ is specified by a random hash function $h:[n]\rightarrow [r]$ and a random $\sigma\in\{\pm 1\}^n$. For each $i\in [n]$ we set $S_{h(i),i}=\sigma_i$, and every other entry in $S$ is set to zero.
\label{def:countsketch}
\end{definition}
In \cite{nelson2013osnap} it is proved that such a matrix satisfies the $(\epsilon,\delta)$-OSE property as long as $r\geq O(d^2/(\delta\epsilon^2))$, $h$ is $2$-wise independent and $\sigma$ is $4$-wise independent. In this work we treat $h$ and $\sigma$ as fully independent, that is, a CountSketch matrix $S$ is assumed to have a single non-zero element per column, equal to $\pm 1$ with probability $1/2$, at a random position chosen uniformly, and that each column is independently sampled.

\subsection{Contributions and structure of this paper}
{
Our main contributions are the following.
\begin{enumerate}
    \item A novel analysis of parallel algorithms and data structures for the CountSketch transform and its combination with Gaussian random projections.
    \item A detailed analysis of parallel algorithms for the computation of the Gramian of a tall-and-skinny matrix $A$.
    \item Applications of the aforementioned methods in column subset selection, least squares regression and leverage scores computation.
    \item {\tt pylspack}, a multithreaded numerical library implementing the aforementioned methods which outperforms existing state-of-the-art libraries.
\end{enumerate}
The outline of the paper is the following. In Section \ref{sec:sketching} we describe a novel sparse data structure for the CountSketch transform. This data structure has low memory requirements, and can be efficiently constructed and multiplied with a sparse or a dense matrix in parallel. We also show that it can be efficiently combined with Gaussian random projections to obtain smaller sketching sizes, which is a common technique in RandNLA. To achieve low memory requirements, the random projection matrices are computed on the fly and discarded once used, in a streaming fashion. Evidently, alternative approaches should be considered when the sketching matrix needs to be persistent in memory for repeated applications \cite{avron2015community}. For the CountSketch matrix, we derive probabilistic bounds regarding load balancing on a shared memory machine.}
{
In Section \ref{sec:gram} we provide parallel algorithms for operations $(iv)$ and $(v)$,} namely the computation of {the Gramian} $A^\top A$ and the computation of the row norms of the product of two matrices. {At the time of this writing, it appears that among the most prominent numerical libraries for matrix computations, the only method targeting the computation of the Gramian of a sparse matrix and returning the result as a dense matrix is {\tt mkl\_sparse\_?\_syrkd} in Intel\footnote{Intel is a trademark or registered trademark of Intel Corporation or its subsidiaries in the United States and other countries.} MKL, which is closed source and its actual implementation is not available. 
}
In Section \ref{sec:applications} we show how these tools can be used to yield efficient parallel algorithms for column subset selection, leverage scores {computation} and least squares regression.
In Section \ref{sec:software} we describe  {\tt pylspack}, a publicly available  Python package,\footnote{\hyperlink{https://github.com/IBM/pylspack}{https://github.com/IBM/pylspack}} whose core is written in C++ and parallelized with OpenMP, and which is compatible with standard matrix data structures of SciPy and NumPy. The package is easily installable with a single command: 
\begin{equation*}
        \texttt{\$ pip install git+https://github.com/IBM/pylspack}
\end{equation*}
{Numerical experiments with a variety of matrices in Section \ref{sec:experiments} demonstrate} that {\tt pylspack} can 
{significantly outperform existing open-source and vendor libraries, especially for the tall-and-skinny case.} 

\subsection{Related work}

The literature on parallel algorithms and data structures for sketching is rather limited, with {most of} the existing works focusing on a small set of sketching transforms, namely Gaussian, sparse Rademacher and subsampled randomized Hadamard transforms \cite{yang2015implementing,avron2010blendenpik,RokhlinTygert.08,meng2014lsrn,wang2007distributed}, while other state-of-the-art transforms, like OSNAP \cite{nelson2013osnap} or the aforementioned CountSketch,\footnote{CountSketch is in fact an OSNAP with sparsity $s=1$ nonzero per column.} have been overlooked.
Software for these tasks is also scarce. In Python, for example, which is arguably the most popular language for data science to date, the only existing implementation of random projections readily available in standard repositories like Anaconda\footnote{\hyperlink{https://anaconda.org/}{https://anaconda.org/}} or Pypi\footnote{\hyperlink{https://pypi.org/}{https://pypi.org/}}, appears to be the {\tt random\_projection} module of Scikit-learn \cite{scikit-learn}, which is in general well implemented, but it is limited to Gaussian transforms and Sparse Sign transforms described in \cite{achlioptas2001database,li2006very} and it is not multithreaded. 
The open source Skylark library \cite{kolliaslibskylark}, which has been used in several works \cite{avron2015community,avron2016high,avron2017faster}, is a prominent example of software for sketching and related algorithms, with great potential in data science applications. It includes implementations for a very large number of sparse and dense sketching transforms and related algorithms, most of them multithreaded{, however, the main focus of Skylark is distributed memory parallelism using MPI, rather than shared memory parallelism, which is the focus of this work}. {Two limitations of Skylark are that} its Python API is implemented in Python 2, which is officially sunset, and it requires substantial effort to build. In our implementation, the {\tt pylspack} Python package, we do not only focus on performance and memory efficiency, but also ease-of-use and applicability, which are desirable properties for a consumable technology.

\subsection{Notation}

We use the Householder notation, that is, matrices are denoted with capital letters, vectors with small letters, and scalars with Greek letters. For a positive integer $n$, $[n]$ is the set $\{1,2,...,n\}$. 
$A_{i,:}$ and $A_{:,j}$ denote the $i$-th row  and $j$-th column of $A$, respectively, both assumed to be column vectors, and $A_{i,j}$ is the element in row $i$ and column $j$. $A_{:,\mathcal{K}}$ denotes the submatrix of $A$ containing the columns defined by the set $\mathcal{K}$. 
$A^\dagger$ is the pseudoinverse. The 2-norm is the default for both matrices and vectors. 
${\tt nnz}(x)$ is the number of nonzeros of $x$, where $x$ can be either a vector or a matrix. 
We define  ${\tt nnz_2}(A)=\sum_{i=1}^{n}\left({\tt nnz}(A_{i,:})\right)^2$, which will be used in complexity bounds. {We use $\tilde O(k):=O(k\log^ck)$, $\tilde \Omega(k):=\Omega(k/\log^ck)$, for some constant $c$, and $f(k)=\tilde \Theta(k)$ is used if $f(k)=\tilde O(k)$ and $f(k)=\tilde \Omega(k)$ at the same time}.
{$\Pr$}$[\alpha]\in[0,1]$ is the probability of occurrence of an event $\alpha$. $\mathcal{N}$ $(\mu,\sigma)$ is the normal distribution with mean value $\mu$ and standard deviation $\sigma$. Matrices with i.i.d. elements from $\mathcal{N}(0,1)$ are referred to as Gaussian matrices. We denote by {\tt randn} standard normal random variables, by {\tt randi} random integers and by {\tt randb} boolean random variables. {
In the compressed sparse row (CSR) format, a matrix $A$ is encoded as three 1-dimensional arrays, $V\in\mathbb{R}^{{\tt nnz}(A)}$,  $J\in\mathbb{N}_+^{{\tt nnz}(A)}$ and $I\in\mathbb{N}_{+}^{n+1}$, which store the values, column indices (ordered by row index) and row pointers, respectively, and $\mathbb{N}_+$ is the set of positive integers}. In particular, the $k$-th element of $I$ shows the starting position in $J$ and $V$ of the $k$-th row of $A$. Therefore, $I_{k+1}-I_{k}$ is the number of nonzero elements of the $k$-th row. In the algorithms and their analysis we will use the notation $A.V$, $A.J$ and $A.I$ to refer to the corresponding arrays $V$, $J$ and $I$ of the matrix $A$.

\section{Sketching algorithms and data structures \label{sec:sketching}}
We consider parallel algorithms for sketching. We first describe data structures for CountSketch transforms, one of the most important random projection constructions in the RandNLA literature. We define certain target properties for storage requirements and complexities, and then propose specific data structures that satisfy them. We then describe efficient parallel algorithms for the combination of Gaussian and CountSketch transforms. Finally, we also describe a memory efficient parallel algorithm for Gaussian sketching, which is very similar to other existing implementations, but we add it for completeness since it is part of our software library. We note that, even though we focus on CountSketch, it is straightforward to adapt our methods for any OSNAP \cite{nelson2013osnap}  (CountSketch is an OSNAP with $s=1$ nonzero per column).

\subsection{CountSketch}
In order to fully exploit parallelism it is useful to store the CountSketch matrix in memory. To this end, we describe a sparse data structure, suitable for CountSketch matrices. This data structure has low memory requirements and can be efficiently constructed and applied to a given input matrix in parallel. Hereafter we assume that the given input matrix is stored in a row-major storage format, specifically, CSR for sparse matrices and row-major array for dense matrices. The output $SA$ is always stored as a dense row-major array. We justify this choice in Section \ref{sec:countgauss}.

Recall that a CountSketch is a matrix $S$ which has exactly one non-zero element in each column at a random position chosen uniformly, and each nonzero is equal to $\pm 1$ with probability $1/2$ \cite{charikar2002finding,clarkson2013low_jacm,meng2013low,nelson2013osnap}. In order to satisfy the OSE property, $S$ needs to have $\Omega(d^2/\epsilon^2)$ rows \cite{nelson2013osnap,nelson2013sparsity}. An example of a ${6}\times 12$ CountSketch matrix is depicted in Figure \ref{fig:countsketch_example}.
Note that, regardless of the number of rows $r$ of $S$, the number of nonzeros is always equal to $n$. An efficient data structure for $S$ should have the following properties.

\begin{definition}
\label{def:specifications}
An efficient CountSketch data structure should adhere to the following specifications, assuming that there are $p$ processors available:
\begin{enumerate}
    \item \label{requirement:1} It stores no more than $O(n)$ elements.
    \item  It admits algorithms which:
    \begin{enumerate}
        \item \label{requirement:2a}
        can construct the data structure in $O(n/p)$ operations,\footnote{{Operations can be of any type, including integer, logic operations and (pseudo) random number generation.}}
    \item \label{requirement:2b} can compute the product $SA$ in $O({\tt nnz}(A)/p)$ operations,
    \item \label{requirement:2c} can compute the product $GSA$, where $G$ is a $m\times r$ dense matrix, in $O((mdr+{\tt nnz}(A))/p)$ operations.
    \end{enumerate}
    \item These algorithms require low (ideally none):
    \begin{enumerate}
        \item \label{requirement:3a} additional storage,
        \item \label{requirement:3b} communication between processors.
    \end{enumerate}
\end{enumerate}

\end{definition}

\subsubsection{Vector representation}
Since each column of a CountSketch matrix $S$ has exactly one nonzero element, a minimal data structure to represent it is to simply store it as a vector $v$ of size $n$, containing signed integers, one for each column. The $k$-th element of $v$, say $v_k$, implies that $S_{|v_k|,k}={\tt sign}(v_k)$. In Figure \ref{fig:countsketch_vector_example} we show an example of a vector representation of the matrix of Figure \ref{fig:countsketch_example}. 

\begin{figure}[H]
    \centering
    $v=\{+4, +2, -5, -6, +1, +6, -6, +2, +1, -5, +5, +3\} $
    \caption{Vector representation of the $6\times 12$ CountSketch matrix of Figure \ref{fig:countsketch_example}.}
    \label{fig:countsketch_vector_example}
\end{figure}

{The vector representation fulfills all the requirements for $p=1$, but it does not lend itself for parallel algorithms. An additional partitioning by rows can remedy this effect.}

\subsubsection{Partitioning}
While the vector representation is simple and has low memory requirements, it is not convenient for computing the products $SA$ and $GSA$ in parallel. A row-major storage, which would allow the independent computation of each row of $SA$, thus eliminating write conflicts, would be preferable. To achieve both low memory requirements as well as efficient parallel computation of $SA$ { and $GSA$, we use partitioning, which is a standard way of performing parallel matrix computations. A straightforward partitioning strategy is to use a dense 2-d array, where each element of the array is a pointer to a sparse matrix data structure storing the corresponding submatrix of $S$. We formalize this in the following \textit{Block CountSketch} (BCS) data structure. As we will show later, this simple data structure satisfies all the required specifications. An example is illustrated in Figure \ref{fig:countsketch_example}.
\begin{definition}
\label{def:block_cs}
Let $S$ be a CountSketch matrix $S$ of size $r\times n$, which is partitioned in $b_r\times b_c$ blocks of size $n_r\times n_c$ each.\footnote{{Without loss of generality, we assume that $n_r$ divides $r$ exactly and $n_c$ divides $n$ exactly, that is, $r=n_r\times b_r$ and $n=n_c\times b_c$.}} A Block CountSketch data structure is a 2-d array of $b_r\times b_c$ abstract data types $B_{i,j}$, $i\in[b_r],j\in[b_c]$. Each $B_{i,j}$ represents a submatrix $S_{i_1:i_2,j_1:j_2}$, where $i_1=(i-1)\times n_r + 1$, $i_2=\min\{i\times n_r, r\}$, $j_1=(j-1)\times n_c + 1$, $j_2=\min\{j\times n_c, n\}$.
\end{definition}}
\begin{figure}[htb]
    \centering
    {\small
    {
    $
    \left(
    \begin{array}{c c c c | c c c c | c c c c}
       0  &  0  &  0 &  0  & -1  &  0  &  0  &  0  & +1  &  0  &  0  &  0 \\
       0  & +1  &  0 &  0  &  0  &  0  &  0  & +1  &  0  &  0  &  0  &  0 \\\hline
       0  &  0  &  0 &  0  &  0  &  0  &  0  &  0  &  0  &  0  &  0  &  +1 \\
      +1  &  0  &  0 &  0  &  0  &  0  &  0  &  0  &  0  &  0  &  0  &  0 \\\hline 
       0  &  0  & -1 &  0  &  0  &  0  &  0  &  0  &  0  & -1  & +1  &  0 \\
       0  &  0  &  0 & -1  &  0  & +1  & -1  &  0  &  0  &  0  &  0  &  0 \\ 
    \end{array}
    \right)
    $
    }
    }
    \caption{A $r\times n$ CountSketch matrix $S$, with $r=6$ and $n=12$, partitioned in $b_r\times b_c=3\times 3$ blocks where each block has size $n_r\times n_c=4\times 2$ (lines denote the partitioning).}
    \label{fig:countsketch_example}
\end{figure}
{
The choice of the data structure to represent submatrices is discussed in detail in Section \ref{sec:submatrices_data_structure}.
}

\subsection{Block CountSketch Algorithms}
{Having a partitioned matrix, performing operations in parallel is rather straightforward. The matrix construction is parallelized across the columns, since the columns are independent but there are dependencies within each column. The multiplication $SA$ is parallelized across rows, since $A$ is assumed to be stored in a row-major format (either sparse CSR or dense row-major). The complexities depend on the actual data structure that will be used to store the blocks of $S$.}
\subsubsection{Construction}
Algorithm \ref{alg:bcs_populate} describes the procedure for populating an empty Block CountSketch instance. The input arguments are integers $r$, $n$, $n_r$, $n_c$, denoting the number of rows, columns, rows-per-row-block and columns-per-column-block of the Block CountSketch matrix, respectively. The complexity is parametrized by the choice of $\mathcal{D}$, which is the data structure type used to store the submatrices (e.g. COO, CSR, CSC, other). Algorithm \ref{alg:bcs_populate} performs $O\left(T_{\tt insert}^{\mathcal{D}}\times (n/p)\right)$ operations, where $T_{\tt insert}^\mathcal{D}$  denotes the complexity of inserting an element in an instance of type $\mathcal{D}$. In order to satisfy Requirement (\ref{requirement:2a}), it must hold that $T_{\tt insert}^\mathcal{D}=O(1)$. 

\begin{algorithm}[htb]
\caption{Block CountSketch: Parallel construction.}
\label{alg:bcs_populate}
\begin{algorithmic}[1]
    \Require Integers $r$, $n$, $n_r$, $n_c$, number of CountSketch rows, columns, rows-per-row-block and columns-per-column-block respectively. $\mathcal{D}$, data structure type to represent submatrices (e.g. COO, CSR, CSC).
    \Ensure Populated Block CS instance of a $r\times n$ CountSketch.
    \State Set $b_r=\lceil \frac{r}{n_r} \rceil,$ $b_c=\lceil \frac{n}{n_c} \rceil$.
      \For{$i=1,...,b_r$} \Comment{parallel for}
          \For{$j=1,...,b_c$}
              \State Instantiate an empty data structure $B_{i,j}$.
          \EndFor
      \EndFor
      \For{$t=1,...,b_c$} \Comment{parallel for}
          \For{$j=(t-1)\times n_c+1,...,\min\{t\times n_c,n\}$}
            \State Generate a row index $i\in[r]$ uniformly at random. \Comment{$O(1)$}
            \State Generate a random value $S_{i,j}\in\{\pm 1\}$ with probability $1/2$. \Comment{$O(1)$}
            \State Find the corresponding block to insert the new element $S_{i,j}$ in the appropriate format. \Comment{$O(T_{\tt insert}^\mathcal{D})$}
          \EndFor
      \EndFor
\end{algorithmic}
\end{algorithm}

\subsubsection{Computing the product $SA$.}
Algorithm \ref{alg:bcs_multiply} describes a procedure for the parallel computation of the product $SA$.
At this point we make the assumption that the product $SA$ will be dense. This is because $SA$ typically consists of $O(d^2)$ rows, in order to satisfy the OSE property \cite{nelson2013osnap,meng2013low}, where each row is constructed as a linear combination of $O(n/d^2)$ rows of $A$ on average. For the tall-and-skinny case, we can assume that $n\gg d^2$, and therefore it is justified to treat and store $SA$ as a dense matrix. 
\begin{algorithm}[htb]
\caption{Block CS: Parallel computation of the product $SA$.}
\label{alg:bcs_multiply}
\begin{algorithmic}[1]
    \Require Matrix $A\in \mathbb{R}^{n\times d}$, populated Block CS instance of a $r\times n$ CountSketch $S$, where each submatrix is stored as an instance of a data structure $\mathcal{D}$.
    \Ensure $C=SA$, the resulting matrix stored as a dense row-major array.
    \State Instantiate $C\leftarrow 0$. \Comment{in parallel}
    \For{$i=1,...,b_r$}\Comment{parallel for}
        \State Set $\mathcal{I}=\left[(i-1)\times n_r+1 ,..., \min\{(i+1)\times n_r, r\}\right]$.
        \For{$j=1,...,b_c$}
            \State Set $\mathcal{J}=\left[(j-1)\times n_c+1,...,\min\{(j+1)\times n_c, n\}\right]$.
            \State Update $C_{\mathcal{I},:} \leftarrow C_{\mathcal{I},:}+S_{\mathcal{I},\mathcal{J}}A_{\mathcal{J},:}$ \Comment{$T_{\tt MM}^\mathcal{D}(S_{\mathcal{I},\mathcal{J}},A_{\mathcal{J},:})$}
        \EndFor
    \EndFor
\end{algorithmic}
\end{algorithm}
Let $T^{\mathcal{D}}_{\tt MM}(S,A)$ denote the complexity of multiplying $S$, stored as a data structure of type $\mathcal{D}$, with a CSR input matrix $A$. The (sequential) complexity of Algorithm \ref{alg:bcs_multiply} is $O\left(\sum_{i=1}^{b_r}\sum_{j=1}^{b_c}T_{\tt MM}^\mathcal{D}(S_{\mathcal{I},\mathcal{J}},A_{\mathcal{J},:})\right)$. Since $S$ is such that each row of $A$ will be used exactly once in the final product, then, as long as the chosen data structure does not perform redundant operations, the complexity is simply $O({\tt nnz}(A))$. It is not straightforward how to derive upper bounds for the parallel complexity using $p\leq b_r$ processors, but due to the randomized nature of $S$ we show that it is possible to obtain probabilistic guarantees in Section \ref{sec:probabilistic_bounds}.

\subsubsection{Computing the product $GSA$ (CountGauss)\label{sec:countgauss}}
The combination of CountSketch and Gaussian matrices has been mentioned in \cite{clarkson2013low_jacm,woodruff2014sketching} and has been extensively studied in \cite{kapralov2016fake} in the context of Support Vector Machines and Nonnegative Matrix Factorizations, where it was first referred to as the ``CountGauss'' transform. It has also been studied in the context of least squares, column selection and leverage scores computation in \cite{sobczyk2021estimating}. Building upon the previous algorithms, we describe an implementation for computing the product $GSA$ where $G$ is a Gaussian matrix and $S$ is a CountSketch.
Parallelization is achieved within each individual step of the algorithm{, which is listed in Algorithm \ref{alg:countgauss_multiply}. To reduce memory requirements for temporary storage, the computation is performed in batches. This does not affect the complexity.}

\begin{algorithm}[htb]
\caption{Block CS: {Batched} parallel computation of the product $GSA$ (CountGauss).}
\label{alg:countgauss_multiply}
\begin{algorithmic}[1]
\Require Matrix $A\in \mathbb{R}^{n\times d}$. Integers $r$, $n$, $n_r$, $n_c$, number of CountSketch rows, columns, rows-per-row-block and columns-per-column-block respectively. $\mathcal{D}$, data structure type to represent submatrices (e.g. COO, CSR, CSC). Batch size $b$.
\Ensure $C=GSA$, where $G$ is a $m\times r$ Gaussian matrix and $S$ is a $r\times n$ CountSketch.
\State Construct a Block CS instance of a $r\times n$ CountSketch $S$ using Algorithm \ref{alg:bcs_populate} with parameters $r,n,n_r,n_c$ and $\mathcal{D}$.
\State Set $n_{\text{batch}}=\lceil\frac{r}{b}\rceil$.
\State Allocate two auxilliary matrices $G\in\mathbb{R}^{m\times b}$ and $B\in\mathbb{R}^{b\times d}$.
\For{$i=1,...,n_{\text{batch}}$}
    \State Let $S_i=S_{(i-1)b+1:\min\{ib,r\},:}$
    \State Compute $B=S_iA$ with Algorithm \ref{alg:bcs_multiply}.
    \State Set new i.i.d elements in $G$ from $\mathcal{N}(0,1)$.\Comment{in parallel}
    \State Update $C = C+GB$. \Comment{with standard BLAS xGEMM}
\EndFor
\State Free $G$ and $B$.
\end{algorithmic}
\end{algorithm}

\subsection{Choosing a data structure for the submatrices\label{sec:submatrices_data_structure}}
{It remains to choose an appropriate data structure to store the submatrices of $S$. We opt to store submatrices in the well-known coordinate format (COO). We do not assume any ordering, therefore constant time insertion can be achieved by appending new elements at the end. 
Moreover, in the unordered COO format the product $SA$ for some input matrix $A$ can be computed in $O({\tt nnz}(A))$ by iterating over the nonzeros $(i,j,v)$ of $S$, multiplying the $j$-th row of $A$ by $v$ and updating the corresponding columns of the $i$-th row of the resulting dense array. 
Hereafter, all references to the COO format pertain to the unordered case.}

{Of interest is the special case of Block CountSketch where $b_r=r$, that is, when the partitioning takes place only column-wise. Since each block of $S$ consists of only one single row, then it can be represented as a sparse vector: it is simply a list of signed integers, each integer denoting the column index of the nonzero element, while its sign denotes whether this nonzero element is equal to $+1$ or $-1$. 
We will refer to this special case as \textit{Block Column CountSketch} (BCCS). 
The complexity of insertion and matrix multiplication for BCCS is the same as COO.
A difference between COO and BCCS is the amount of memory required to store the entire Block CountSketch matrix. 
For BCCS, we only store one signed index for each nonzero element, totaling $n$ elements. In addition, $r\times b_n$ pointers to the submatrices are stored. In summary, the memory requirements are as follows:
\begin{itemize}
    \item COO: $2n$ indices + $b_r\times b_n$ pointers,
    \item BCCS: $n$ indices + $r\times b_n$ pointers.
\end{itemize}
$n$ additional bits are required to store the signs in both representations.
}

\subsection{Probabilistic bounds for processor workload\label{sec:probabilistic_bounds}}
{
Due to the randomized nature of the CountSketch matrix, we can derive upfront probabilistic bounds on the workload distribution among processors.
Let $S$ be a $r\times n$ CountSketch matrix and $A\in\mathbb{R}^{n\times d}$. Let $w_i={\tt nnz}(A_{i,:})$. Recall that the product $SA$ is computed in $O({\tt nnz}(A))=O(\sum_{i=1}^n w_i)$. Assume that we have $p$ processors, and the multiplication is performed in parallel as follows: each processor computes $B=r/p$ randomly chosen rows of $SA$.  Each row $j$ of $SA$, $j\in[r]$, is simply the product of $S_{j,:}A$. Let $Q\subset[n]$, be the set of indices of nonzero columns of $S_{j,:}$. Clearly, the product $S_{j,:}A$ is computed in $O(\sum_{k\in Q}w_k)$. Since $S$ is a random matrix, we can try to get probabilistic bounds on the ``workload'' of each processor. Let $Y=\sum_{k\in Q}w_k$ be the random variable which counts the number of nonzeros of $A$ that ``contribute'' to the product $S_{j,:}A$. This variable can also be rewritten as $Y=\sum_{i=1}^n \epsilon_iw_i$, where $\epsilon_i$ is one if $w_j$ was chosen to contribute in $S_{j,:}A$ and zero otherwise. But $\epsilon_i$ is simply a Bernoulli random variable with parameter $q=1/r$ since each column of $S$ has only one nonzero in a position which is chosen uniformly at random from the set $[r]$. The $\epsilon_i$'s are also independent since each column is sampled independently. Generalizing for a fixed set of $B$ rows instead of simply one row we have the following.
\begin{lemma}
\label{lem:expectation}
Let $S$ be a $r\times n$ CountSketch and $A$ a fixed $n\times d$ input matrix. Let $Q\subset[r]$, $|Q|=B<r$ be a subset of $B$ rows of $S$, chosen uniformly at random without replacement. For $i\in[n]$, let $w_i\leq d$, $\sum_{i=1}^nw_i={\tt nnz}(A)$, and $\epsilon_1,...,\epsilon_n$ be independent Bernoulli random variables with parameter $0<q=B/r=1/p< 1$, that is, $\epsilon_i=1$ with probability $q$ and $\epsilon_i=0$ with probability $1-q$. Let $Y=\sum_{i=1}^n\epsilon_iw_i$. The following hold:
\begin{align*} 
\mu &=\mathbb{E}[Y]=q{\tt nnz}(A),\\
\sigma^2 &= \text{Var}(Y)=q(1-q){\tt nnz_2}(A).
\end{align*}
\begin{proof}
For the expectation we have
\begin{align*}
    \mu=\mathbb{E}[Y] = \mathbb{E}\left[\sum_{i=1}^n w_i \epsilon_i\right] = \sum_{i=1}^n w_i\mathbb{E}[\epsilon_i] = q\sum_{i=1}^nw_i = q {\tt nnz}(A).
\end{align*}
Similarly, we can compute the variance. First we need to compute $\mathbb{E}[Y^2]$. We have
\begin{align*}
    \mathbb{E}Y^2
    &=\mathbb{E}\left[(\sum_{i=1}^n w_i\epsilon_i)^2\right] 
    = \mathbb{E}\left[(\sum_{i=1}^nw_i^2\epsilon_i^2) +2\sum_{i\neq k}w_iw_k\epsilon_i\epsilon_{k}\right] = q\sum_{i=1}^nw_i^2 + 2q^2\sum_{i\neq k}w_iw_k.
\end{align*}
We can then compute
\begin{align*}
    \text{Var}(Y)=\mathbb{E}[Y^2]-(\mathbb{E}[Y])^2 = q(1-q)\sum_{i=1}^nw_i^2=q(1-q){\tt nnz_2}(A).
\end{align*}
\end{proof}
\end{lemma}
We can then use Chebyshev's inequality to get a first tail bound. 
\begin{corollary}
Let $A$, $q$ and $Y$ be as in Lemma \ref{lem:expectation}. It holds that
\begin{align*}\Pr\left[
        |Y-\mathbb{E}Y|
        \geq 
        \sqrt{nq(1-q){\tt nnz_2}(A)}
    \right]
    \leq 
    \frac{1}{n}.\end{align*}
\begin{proof}
The result follows by simply applying Chebyshev's inequality on the random variable $Y$ and using Lemma \ref{lem:expectation}. This gives \begin{align*}\Pr\left[|Y-\mu|\geq t\right]\leq \frac{\sigma^2}{t^2}.\end{align*}
The result follows by substituting $t=\sigma\sqrt{n}=\sqrt{nq(1-q){\tt nnz_2}(A)}$.
\end{proof}
\end{corollary}
As it is known, the inequalities of Chebyshev and Markov are generally tight. We can say more by noting that the random variables are bounded in magnitude.
Therefore, we can use Hoeffding's inequality to get a tighter tail bound. For proofs of Theorem \ref{thm:hoeffding} see the corresponding references.
\begin{theorem}[Hoeffding's Inequality \cite{bandeira2020mathematics,vershynin2018high}]
\label{thm:hoeffding}
Let $Z_1,...,Z_n$ be independent, bounded and centered random variables, that is $|Z_i|\leq \alpha_i$ and $\mathbb{E}[Z_i]=0$. Then,
\begin{align*}\Pr\left[
    \left|
        \sum_{i=1}^n Z_i
    \right|
    >t
\right] 
\leq 
2\exp\left(
    -\frac{t^2}
    {2\sum_{i=1}^n\alpha_i^2}
\right).\end{align*}
\end{theorem}
To apply Hoeffding's inequality, it suffices to derive the appropriate centered random variables. To that end, let $Z_i=Y_i-\mathbb{E}Y_i=\epsilon_iw_i-qw_i=(\epsilon_i-q)w_i$. Clearly, $Z_i\in[-qw_i,(1-q)w_i]\Rightarrow|Z_i|\leq (1-q)w_i=\alpha_i$ if we assume that $q\leq 1/2$.
\begin{lemma}
\label{lemma:concentration}
Let $Y_i$, $A$ and $S$ be the same as in Lemma \ref{lem:expectation} and $Z_i=Y_i-\mathbb{E}Y_i$. The following hold
\begin{align*}
    \Pr\left[
        \left|
            \sum_{i=1}^n
            Z_i
        \right|
        >
        (1-q)\sqrt{2{\tt nnz_2}(A)\log n}.
    \right]
    &\leq 
    \frac{2}{n},
    \\
    \Pr\left[
        \left|
            \sum_{i=1}^n
            Z_i
        \right|
        >
         \mu\sqrt{2\left(\tfrac{1-q}{q}\right)^2\log n}.
    \right]
    &\leq 
    \frac{2}{n}.
\end{align*}
\begin{proof}
Since $Z_i$ are independent, centered, and bounded, we can directly use Hoeffding's inequality. This gives
\begin{align*}\Pr\left[
    \left|
        \sum_{i=1}^n Z_i
    \right|
    >t
\right] 
\leq 
2\exp\left(
    -\frac{t^2}
    {
        2(1-q)^2{\tt nnz_2}(A)
    }
\right).\end{align*}
Setting $t=(1-q)\sqrt{2{\tt nnz_2}(A)\log n}$ gives the first inequality. For the second inequality, we set $t=\delta\mu$, where $\mu=q{\tt nnz}(A)$ and $\delta>0$. This gives 
\begin{align*}\Pr\left[
    \left|
        \sum_{i=1}^n Z_i
    \right|
    >
    \delta \mu
\right] 
\leq 
2\exp\left(
    -\frac{\delta^2\mu^2}
    {
        2(1-q)^2{\tt nnz_2}(A)
    }
\right)
=
2\exp\left(
    -\frac{\delta^2q^2{\tt nnz}^2(A)}
    {
        2(1-q)^2{\tt nnz_2}(A)
    }
\right)
.\end{align*}
Note that $1\leq \frac{{\tt nnz}^2(A)}{{\tt nnz_2}(A)}\leq n$, since these two quantities are simply the norms $\|w\|^2_2$ and $\|w\|_1^2$ where $w$ is a $n$-dimensional vector containing all $w_i$. Moreover, $q/(1-q)=\frac{B/r}{(1-B/r)}=\frac{B}{r-B}\in[1/(r-1),1]$ for $B\in[1,r/2]$. Therefore the failure probability can be further bounded as
\begin{align*}
2\exp\left(
    -\frac{\delta^2q^2{\tt nnz}^2(A)}
    {
        2(1-q)^2{\tt nnz_2}(A)
    }
\right)
\leq
2\exp\left(
    -\frac{\delta^2q^2}
    {
        2(1-q)^2
    }
\right).
\end{align*}
Setting $\delta=\frac{1-q}{q}\sqrt{2\log n}$ concludes the proof for the last inequality.
\end{proof}
\end{lemma}
This implies that, for a fixed subset of $[r]$ with size $B$, large fluctuations in the corresponding workload have small probability to occur. We have finally left to analyze the behaviour over all $p$ processors. The analysis is finalized by taking a union bound over $p$ sets, with $q=B=r/p$ rows in each set. We summarize in the following lemma.
\begin{lemma}
\label{lemma:balancing}
Let $A,S$ be the same as in Lemma \ref{lem:expectation}. If we use $p$ processors to perform the multiplication $SA$, by assigning $B=r/p$ distinct rows of $S$ on each processor, then the following hold:
\begin{enumerate}
    \item Each processor will execute $O({\tt nnz}(A)/p)$ operations in expectation,
    \item the bounds of Lemma \ref{lemma:concentration} hold for all processors at the same time with probability at least $1-\tfrac{2p}{n}$.
\end{enumerate}
\begin{proof}
The expectation comes directly from Lemma \ref{lem:expectation}. For the tail, we simply take the union bound of success probabilities of $p$ distinct fixed sets.
\end{proof}
\end{lemma}
We conclude the analysis of CountSketch data structures with the following theorem.
\begin{theorem}
\label{thm:block_countsketch_satisfy_conditions}
Given a sparse tall-and-skinny matrix $A$, the Block-COO-CountSketch and Block-Column-CountSketch data structures satisfy all conditions of Definition \ref{def:specifications} deterministically, except Condition (\ref{requirement:2b}), which is satisfied in expectation and it is well centered with high probability based on Lemma \ref{lemma:balancing}.
\end{theorem}
}

\subsection{Gaussian sketching}
Unlike other sketching transforms, Gaussian projections have been well studied not only in theory but also in practice, most notably for computing randomized least squares preconditioners \cite{RokhlinTygert.08,meng2014lsrn}. The description that we provide here is very similar to the aforementioned works. 
If $A$ is dense we can directly use BLAS routines to compute the product. If $A$ is sparse in CSR format, one can take advantage of the sparsity as described in Algorithm \ref{alg:gaussian_sketch}. To parallelize, each process is assigned a block of rows of $G$, say $G_p$, and computes the entire product $G_p A$. In this manner, there are no write conflicts and no need for synchronization during execution, while each process executes the same number of operations. However, as well noted in \cite{meng2014lsrn}, it is crucial to use a fast algorithm for generating Gaussian random variables, otherwise the total performance might be overwhelmed by this operation. 

\begin{algorithm}[htb]
\caption{Parallel sketch of a CSR matrix with a Gaussian matrix.}
\label{alg:gaussian_sketch}
\begin{algorithmic}[1]
    \Require CSR Matrix $A\in \mathbb{R}^{n\times d}$, integer $m$, number of processors $p$.
    \Ensure $C=GA$, where $G$ is a $m\times n$ Gaussian matrix.
    \State Set $b=\left\lceil\frac{m}{p}\right\rceil$.
    \For{$t=1,...,p$} \Comment{parallel for}
      \State Set $b_t=\min\{b,n-(t-1)b\}$.
      \State Allocate vector $g$ of size $b_t$.
      \For{$k=1,...,n$}
          \State Set new i.i.d elements in $g$ from $\mathcal{N}(0,1)$.
          \For{$h=A.I_k,...,A.I_{k+1}$}
            \For{$i=tb+1,...,tb+b_t$}
              \State Denote $j=A.J_h$.
              \State $C_{i,j} = C_{i,j} + g_i\times A.V_h $.
            \EndFor
          \EndFor
      \EndFor
      \State Free $g$.
    \EndFor
\end{algorithmic}
\end{algorithm}

\section{Gram matrix and Euclidean row norms\label{sec:gram}}

{Given a sparse matrix $A\in\mathbb{R}^{n\times d}$, the computation of the Gram matrix $A^\top A$ is a fundamental operation in various applications. This can be performed with standard general sparse matrix multiplication algorithms, based on Gustavson's algorithm
\cite{gustavson1978two} or more recent proposals
\cite{azad2016exploiting,buluc2008representation}; see also \cite[Chapters 13 and 14]{kepner2011graph} for a review. However, most works focus on the general case $C=AB$ where the inputs $A,B$ are distinct sparse matrices,\footnote{{Some recent works \cite{arrigoni2021efficiently,dumas2020fast} are focused on multiplying $A$ by its transpose, albeit for dense matrices.}} and the resulting matrix is also sparse, and therefore they consider efficient access patterns and sparsity structures for $C$. Here, instead, we consider the special case of the Gramian $A^\top A$ when $A$ is tall-and-skinny. The task, in general, is not easy, and this has been recognized in several implementations and software libraries for sparse matrix computations \cite{saad1990sparskit,duff2002overview}.  From a theoretical perspective, the fastest known sequential algorithms are based on the concept of fast matrix multiplication. As it is known, two general rectangular dense matrices $A\in\mathbb{R}^{m\times k}$ and $B\in\mathbb{R}^{k\times n}$ can be multiplied in $O(mnk^{\omega-2})$, where $\omega<2.37286$ is the best known bound for the exponent of square matrix multiplication \cite{alman2021refined}. Analogous results exist for explicitly rectangular matrix multiplication \cite{gall2018improved} and for sparse matrix multiplication \cite{yuster2005fast}. Outer-product based algorithms that are considered in this work might be asymptotically ``slower'', however, they are simple to analyze and work well in practice.
}

{
A key observation is that when $A$ is sufficiently tall, then the resulting Gram matrix will most likely be dense.  This assumption greatly simplifies the complexity of efficient algorithms. In fact, there is a remarkably simple algorithm to perform this operation which is  {optimal} in the sense that it performs $\Theta(R)$ scalar multiplications, where $R$ is the number of {required} scalar multiplications to be performed by any standard inner-product or outer-product based algorithm (see also \cite[Section 1.4.3]{kepner2011graph}). We list this algorithm in Algorithm \ref{alg:gram_serial}. 
\begin{algorithm}[htb]
\caption{CSR Gram.}
\label{alg:gram_serial}
\begin{algorithmic}[1]
\Require CSR Matrix $A\in \mathbb{R}^{n\times d}$, row-major matrix $B\in\mathbb{R}^{d\times d}$.
\Ensure $B\leftarrow A^\top A$.
\For{$i=1,...,n$}
    \For{$k=A.I_i,...,A.I_{i+1}$}
        \For{$j=A.I_i,...,A.I_{i+1}$}
            \State Denote $i'=A.J_{k}$ and $j'=A.J_j$.
            \State $B_{i',j'} = B_{i', j'} + A.V_k \times A.V_j$.
        \EndFor
    \EndFor
\EndFor
\end{algorithmic}
\end{algorithm}
Notably, at the time of writing, none of the following scientific software libraries, all of which include sparse matrix components, provide function calls targeting this specific operation: Eigen \cite{eigenweb}, SPARSKIT \cite{saad1990sparskit}, CombBLAS \cite{bulucc2011combinatorial}, SciPy \cite{virtanen2020scipy}, IBM\textregistered \  ESSL \cite{essl},\footnote{IBM and the IBM logo are trademarks of International Business Machines Corporation, registered in many jurisdictions worldwide. Other product and service names might be trademarks of IBM or other companies. Acurrent list of IBM trademarks is available on ibm.com/trademark.} Ginkgo \cite{anzt2020ginkgo}, CSparse \cite{davis2006direct}. It appears that the only available implementations of this operation are the Intel MKL \cite{mkl} {\tt mkl\_sparse\_?\_spmmd} and {\tt mkl\_sparse\_?\_syrkd} methods, which are  closed-source and the actual implementation used is not available. Algorithm \ref{alg:gram_serial} is simple and optimal in terms of operations. However, due to the irregular sparsity patterns of the outer products formulation, it is not straightforward to implement it in parallel. 
}

\subsection{Parallel Gram 1: A memory efficient algorithm}
{We describe a memory-efficient parallel variant of Algorithm \ref{alg:gram_serial}, which requires no additional memory for intermediate results and no communication between processors. The Algorithm works as follows: each processor $t$ is assigned a predifined range $\mathcal{T}=[i_1,i_2], 1\leq i_1\leq i_2\leq d$, of rows of $A^\top$ and computes the product $B_{\mathcal{T},:}\leftarrow (A^\top)_{\mathcal{T},:}A$. However, since the matrix $A$ is stored in CSR, and therefore $A^\top$ is stored in CSC, the processors do not have constant time access to the rows of $A^\top$. Each processor needs to filter out the rows that it needs to discard. That is, for each row $i\in[n]$ of $A$, each processor must check which column indices of that row belong in the range $\mathcal{T}$. A sequential search runs in $O({\tt nnz}(A_{i,:}))$, summing up to $O(\sum_{i=1}^n{\tt nnz}(A_{i,:}))=O({\tt nnz}(A))$. However, if column indices of each row are sorted in an ascending or descending order, which is the case for CSR, then search can be performed in a binary fashion to yield a total of $O(\sum_{i=1}^n\max\{1,\log({\tt nnz}(A_{i,:}))\})=\Omega(n)$ and $O(n\log d)$. Moreover, unless the distribution of nonzeros of $A$ among the rows and the columns is highly imbalanced, or more specifically, if each row has $\tilde \Theta ({\tt nnz}(A)/n))$ nonzeros and each column has $\tilde \Theta ({\tt nnz}(A)/d)$ nonzeros, then the time complexity of the actual multiplication operation is $\tilde \Theta ({\tt nnz_2}(A)/p)$. This means that the time complexity of the Algorithm is $\tilde \Theta (n+{\tt nnz_2}(A)/p)$, which is close to the desired $O({\tt nnz_2}(A)/p)$. 
As shown in the experimental evaluation, the algorithm scales adequately even for matrices with $\sim 4\%$ non-zeros.}

\begin{algorithm}[htb]
\caption{Parallel CSR Gram with no additional memory requirements and no communication.}
\label{alg:gram_parallel_memory_effcient}
\begin{algorithmic}[1]
\Require CSR Matrix $A\in \mathbb{R}^{n\times d}$, row-major matrix $B\in\mathbb{R}^{d\times d}$, scalars $\alpha,\beta$, number of processors $p$.
\Ensure $B\leftarrow \alpha A^\top A+\beta B$.
\If{$\beta \neq 1$}
    \For{$i=1,...,d$} \Comment{parallel for}
        \For{$i=1,...,d$}
            \State $B_{i,j} = \beta B_{i,j}$.
        \EndFor
    \EndFor
\EndIf
\If{$\alpha\neq 0$}
    \State Set $b=\left\lceil\frac{n}{p}\right\rceil$.
    \For{each processor $t=1,...,p$} \Comment{parallel for}
        \For{$i=1,...,n$}
            \For{$k=A.I_i,...,A.I_{i+1}$}
                \State $i'=A.J_k$
                \If{$i'\in[(t-1)b+1,\min\{tb,n\}]$}
                    \State Set $\gamma=\alpha\times A.V_k$
                    \For{$h=I_i,...,I_{i+1}$}
                        \State $B_{i', A.J_h} = B_{i', A.J_h} + \gamma \times A.V_h$.
                    \EndFor            
                \EndIf
            \EndFor
        \EndFor
    \EndFor
\EndIf
\end{algorithmic}
\end{algorithm}

\subsection{Parallel Gram 2: A faster algorithm \label{sec:parallel_gram}}
{
Faster algorithms can be derived if additional memory is available to store temporary results. One such algorithm can procceed by splitting the matrix $A$ in row blocks, assigning each processor to compute the Gram matrix of one row block, storing it locally in a dense matrix, and ultimately summing all the partial results. This algorithm requires $pd^2$ additional storage for intermediate results. A similar analysis to the one of the previous section, shows that for any matrix with $\tilde O({\tt nnz}(A)/n)$ nonzeros per row, each processor in Algorithm \ref{alg:gram} will compute the corresponding Gram matrix of the rows that it was assigned in $\tilde O({\tt nnz_2}(A)/p)$ steps. Thereafter, the partial results will be summed in $p\times(d^2/p)=d^2$ steps, therefore, the total time complexity of the algorithm is $\tilde O({\tt nnz_2}(A)/p + d^2)$. For the tall-and-skinny case, i.e. when $n\gg d^2$, this algorithm has a much smaller time complexity than Algorithm \ref{alg:gram_parallel_memory_effcient}. In future work, it would be of interest to also take I/O  into account, in the spirit of recent research \cite{kwasniewski2019red,kwasniewski2021parallel}.
}

\begin{algorithm}[htb]
\caption{Parallel CSR Gram based on row partitioning.}
\label{alg:gram}
\begin{algorithmic}[1]
\Require CSR Matrix $A\in \mathbb{R}^{n\times d}$, row-major matrix $B\in\mathbb{R}^{d\times d}$, scalars $\alpha,\beta$, number of processors $p$.
\Ensure $B\leftarrow \alpha A^\top A+\beta B$.
\If{{$\beta\neq 1$}}
    \For{$i=1,...,d$} \Comment{parallel for}
        \For{{$j=1,...,d$}}
            \State $B_{i,j} = \beta B_{i,j}$.
        \EndFor
    \EndFor
\EndIf
\If{{$\alpha\neq 0$}}
    \State Set $b=\left\lceil\frac{n}{p}\right\rceil$.
    \State Set $c=\left\lceil\frac{d}{p}\right\rceil$.
    \For{each processor $t=1,...,p$} \Comment{parallel for}
        \State Allocate $B^t\in\mathbb{R}^{d\times d}$ and set each element to $0$.
        \State Denote $\mathcal{T}=[(t-1)b+1,\min\{tb,n\}]$.
        \State Denote $A^t=A_{\mathcal{T},:}$
        \State Compute $B^t \leftarrow (A^t)^\top A^t$ using Algorithm \ref{alg:gram_serial}.
        \State \# Barrier: Synchronize all processors \Comment{Reduction Step}
        \State Denote $\mathcal{R}=[(t-1)c+1,\min\{tc,d\}]$.
        \For{$k=1,...,p$}
            \State Update $B_{\mathcal{R},:}\leftarrow B_{\mathcal{R},:} + B^k_{\mathcal{R},:}$        
        \EndFor
    \EndFor
\EndIf
\end{algorithmic}
\end{algorithm}

\subsection{Computing Euclidean row norms}
We {next} describe {an} algorithm for computing the squared row norms of the matrix product ${C=}AB$. 
{As we will see, this is a computational kernel in advanced algorithms for leverage scores computation, with $A$ being the matrix whose leverage scores are sought,} while $B$ is an ``orthogonalizer'' for $A$, {i.e. it is} such that $AB$ {forms an (exact or approximate) orthonormal basis for range$(A)$}.
A naive approach to compute the row norms is to first compute {the product } ${C=}AB$ and then compute the {squared euclidean row norms}. {An alternative approach that has the same complexity but} lower memory requirements is to {avoid precomputing the product and  instead,} iterate over  the rows of $A$, computing $\|(e_i^\top A)B\|$ and storing only the final result. For sparse matrices, there is an even better approach, achieving $O({\tt nnz_2}(A))$ complexity, instead of $O({\tt nnz}(A)d)$; cf. \cite[Lemma 2.1]{sobczyk2021estimating}.

\begin{algorithm}[htb]
\caption{Parallel squared row norms of the product of a CSR matrix with a dense row-major matrix.}
\label{alg:squared_row_norms}
\begin{algorithmic}[1]
\Require CSR Matrix $A\in \mathbb{R}^{n\times d}$, dense row-major matrix $B\in\mathbb{R}^{d\times r}$, scalars $\alpha,\beta$.
\Ensure $x\leftarrow \alpha q + \beta x$,  where $q$ is a vector containing the squared Euclidean norms of the rows of $AB$.
\If{{$\beta\neq 1$}}
    \For{$i=1,...,n$}\Comment{parallel for}
        \State $x_i = \beta x_i$.
    \EndFor
\EndIf
\If{{$\alpha\neq 0$}}
    \State Compute $\tilde B = BB^\top$. \Comment{with standard BLAS xGEMM.}
    \For{$i=1,...,n$}\Comment{parallel for}
        \State Set $\tilde x = 0$.
        \For{$k=A.I_i,...,A.I_{i+1}$}
            \State Set $\gamma = \alpha \times A.V_k$.
            \For{$j=A.I_i,...,A.I_{i+1}$}
                \State Denote $i'=A.J_j$ and $j'=A.J_k$.
                \State $\tilde x = \tilde x + \gamma\times A.V_j \times \tilde B_{i',j'}$.
            \EndFor
        \EndFor
        \State $x_i = x_i + \tilde x$.
    \EndFor
\EndIf
\end{algorithmic}
\end{algorithm}

Algorithm \ref{alg:squared_row_norms} performs this operation for a sparse CSR input matrix and achieves the desired $O({\tt nnz_2}(A))$ number of operations.
There are no write conflicts during execution and therefore the outer loop can be fully parallelized. 
{In our implementation, t}o improve load balancing, we do not restrict a specific list of rows for each process to compute. Instead, we let each process choose a new element of $x$ to compute once it has finished its previous computation.

\section{Applications \label{sec:applications}}
{We next demonstrate some example applications of the proposed methods}. We note that the high level algorithms that we will describe are parallel variants of algorithms that have already been discussed in the literature. We thus omit details related to the proofs of correctness and statistical guarantees and refer the reader to the corresponding publications.

\subsection{Column subset selection}
In Algorithm \ref{alg:cgrrqr} we describe an algorithm to select a subset of $k$ columns of a tall-and-skinny matrix. A practical choice is to use $m\approx2d$ rows for $G$ and $r\approx d^2$ rows for $S$ to achieve decent speed as well as approximation guarantees. We refer to
\cite{cheung2013fast,boutsidis2009improved, avron2013faster,civril2014column,shitov2021column,sobczyk2021estimating}
for more details. Hereafter, by $A_k$ we denote the best rank-$k$ approximation of $A$ in the $2$-norm.

\begin{algorithm}[htb]
\caption{Parallel column subset selection with CountGauss. \cite[Algorithm 4.2]{sobczyk2021estimating}}
\label{alg:cgrrqr}
\begin{algorithmic}[1]
\Require CSR Matrix $A\in \mathbb{R}^{n\times d}$.
\State Compute $B=GSA$ with Algorithm \ref{alg:countgauss_multiply} with $m\approx 2d$ and $r\approx d^2$.
\State Run a pivoted QR on $B$ to obtain a permutation matrix $P$. 
\State Use {the first $k$ columns of} $P$ to select the ``top-$k$'' columns of $A$.
\end{algorithmic}
\end{algorithm}

\subsection{Least squares regression}
In Algorithm \ref{alg:least_squares_gram} we describe a direct least squares solver based on Algorithm \ref{alg:gram}. In this case $A^\top A$ is explicitly formed and its pseudoinverse is used to compute the solution of the least squares problem.

\begin{algorithm}[htb]
\caption{Parallel least squares regression via Gram pseudoinverse.}
\label{alg:least_squares_gram}
\begin{algorithmic}[1]
\Require CSR Matrix $A\in \mathbb{R}^{n\times d}$.
\State Compute $B=A^\top A$ with Algorithm \ref{alg:gram_parallel_memory_effcient}.
\State Compute $V,\Sigma, V^\top = {\tt svd}(B)$. \Comment{Truncate if needed.}
\State Compute $c = A^\top b$.
\State \Return $x=V\Sigma^\dagger V^\top c$.
\end{algorithmic}
\end{algorithm}

In Algorithm \ref{alg:least_squares_approx} we describe a direct least squares solver based on Algorithm \ref{alg:countgauss_multiply}. Such algorithms in the literature are often referred as ``sketch-and-solve''. In this case we first sketch the matrix $\begin{pmatrix}A & b\end{pmatrix}$ to a smaller size using a sketching matrix $S$ which satisfies the OSE property and then directly solve the least squares problem $\arg\min_x\|SAx-Sb\|$. In this case the solution is approximate. For further details see \cite{sarlos2006improved,Drineas2010,nelson2013osnap,woodruff2014sketching}.

\begin{algorithm}[htb]
\caption{Approximate least squares solution with CountGauss.}
\label{alg:least_squares_approx}
\begin{algorithmic}[1]
\Require CSR Matrix $A\in \mathbb{R}^{n\times d}$.
\State Compute $\begin{pmatrix} \tilde A & \tilde b\end{pmatrix}=GS\begin{pmatrix}A & b\end{pmatrix}$ with Algorithm \ref{alg:countgauss_multiply}.
\State Compute $U,\Sigma, V^\top = {\tt svd}(\tilde A)$. \Comment{Truncate if needed}
\State \Return $x = V \Sigma ^\dagger U^\top\tilde b$.
\end{algorithmic}
\end{algorithm}

In Algorithm \ref{alg:least_squares_precondition} we describe an iterative least squares solver, based on building a preconditioner for $A^\top A$ using Algorithm \ref{alg:countgauss_multiply} and then using a preconditioned iterative method to solve the least squares problem. Such algorithms in the literature have been referred as ``sketch-and-precondition''. {$m$ denotes the number of rows of the Gaussian embedding $G$, and $r$ denotes the number of rows of the CountSketch matrix $S$ (note that $r$ is also the number of columns of $G$). By choosing $m=0$, then only the CountSketch transform will be applied on $A$, e.g. only $SA$ will be computed, ignoring $G$.} Practical combinations for $m$ and $r$ are $m\approx 2d$ and $r\approx d^2$ (\cite{meng2014lsrn,sobczyk2021estimating}) or even $m=0$ and $r\approx 2d$ \cite{dahiya2018empirical}; see \cite{RokhlinTygert.08,Drineas2010,avron2010blendenpik,meng2014lsrn,pilanci2016iterative}
for more details on sketching algorithms for least squares regression.

\begin{algorithm}[htb]
\caption{Parallel preconditioned iterative least squares with CountGauss.}
\label{alg:least_squares_precondition}
\begin{algorithmic}[1]
\Require Matrix $A\in \mathbb{R}^{n\times d}$.
\State Compute $B=GSA$ with Algorithm \ref{alg:countgauss_multiply}.
\State Compute $U,\Sigma,V^\top={\tt svd}(B)$. \Comment{Truncate if needed.}
\State Compute $N=V\Sigma^{-1}$.
\State Solve $y^*=\arg\min_y\|ANy-b\|$ with a parallel preconditioned iterative method.
\State \Return $x=Ny^*$.
\end{algorithmic}
\end{algorithm}

\subsection{Leverage scores}
Using the basic algorithms of Sections 2 and 3 we can describe parallel algorithms for leverage scores. For details on accuracy and parameter selection we refer the reader to \cite[\S 5]{sobczyk2021estimating}. Algorithm \ref{alg:ls_direct} computes the exact leverage scores by first computing the Gram matrix and then using its pseudoinverse as an orthogonalizer for $A$. Algorithm \ref{alg:ls_hrn_exact} first selects a subset of columns using Algorithm \ref{alg:cgrrqr}. Algorithms \ref{alg:ls_approximate_drineas} and \ref{alg:ls_hrn_approx} return approximate solutions and are suited for dense matrices.

\begin{algorithm}[htb]
\caption{Parallel leverage scores via Gram pseudoinverse.}
\label{alg:ls_direct}
\begin{algorithmic}[1]
	\Require Matrix $A\in \mathbb{R}^{n\times d}$.
	\Ensure leverage scores $\theta_i$ of $A$ over its dominant $k$-subspace.
	\State Compute $B= A^\top A$ using Algorithm \ref{alg:gram}.
	\State Compute $V,\Sigma^2, V^\top = {\tt svd}(B)$. \Comment{Truncate if needed.}
	\State Compute $\theta_i=\|e_i^\top AV\Sigma^{-1}\|^2,i\in[n]$ using Algorithm \ref{alg:squared_row_norms}.
	\State \Return $\theta_i$. 
\end{algorithmic}
\end{algorithm}

\begin{algorithm}[htb]
	\caption{Parallel leverage scores of $A_k$ based on \cite[Algorithm 5.3]{sobczyk2021estimating}}
	\label{alg:ls_hrn_exact}
	\begin{algorithmic}[1]
		\Require matrix $A \in \mathbb{R}^{n \times d}$.
		\Ensure $\tilde\theta_i(A_k)$, {approximate leverage scores of $A_k$ where $k\leq d$ (exact if $k=\text{rank}(A)$).}
        \State Select $k$ columns in $A_{:,\mathcal{K}}$ using Algorithm \ref{alg:cgrrqr}.
		\State Compute $\tilde \theta_i,i\in[n]$ using Algorithm \ref{alg:ls_direct} on $A_{:,\mathcal{K}}$.
		\State \Return $\tilde \theta_i$.
	\end{algorithmic}
\end{algorithm}

\begin{algorithm}[htb]
	\caption{Approximate leverage scores via sketched SVD. Parallel variant of \cite{clarkson2013low_jacm,drineas2012fast,nelson2013osnap}.}
	\label{alg:ls_approximate_drineas}
	\begin{algorithmic}[1]
		\Require matrix $A \in \mathbb{R}^{n \times d}$ with rank$(A)=d$.
		\Ensure $\tilde \theta_i$, approximate leverage scores of $A$.
        \State Construct and populate a $O(\frac{d^2}{\epsilon^2})\times d$ CountSketch $S$.
        \State Compute $\tilde A= SA$ with Algorithm \ref{alg:bcs_multiply}. 
        \State Compute $\tilde U, \tilde \Sigma,\tilde V^\top={\tt svd}(\tilde A)$.
        \State Set $\Pi$ a $d\times O(\log n/\epsilon^2)$ scaled Gaussian matrix.
	    \State Compute $X= \tilde V \tilde \Sigma^{-1}\Pi$. \State Compute $\tilde\theta_i=\|e_i^\top AX\|_2^2, \forall i\in [n]$ using Algorithm \ref{alg:squared_row_norms}.
	    \State \Return $\tilde\theta_i$.
	\end{algorithmic}
\end{algorithm}

\begin{algorithm}[htb]
	\caption{Approximate leverage scores, parallel variant of \cite[Algorithm 5.6]{sobczyk2021estimating}}
	\label{alg:ls_hrn_approx}
	\begin{algorithmic}[1]
		\Require matrix $A \in \mathbb{R}^{n \times d}$.
		\Ensure $\tilde \theta_i(A_k)$, approximate leverage scores of $A_k$, where $k\leq d$.
        \State Select $k$ columns in $A_{:,\mathcal{K}}$ using Algorithm \ref{alg:cgrrqr}.
        \State Compute $\tilde \theta_i$ using Algorithm \ref{alg:ls_approximate_drineas} on $A_{:,\mathcal{K}}$.
	    \State \Return $\tilde \theta_i$.
	\end{algorithmic}
\end{algorithm}

\section{Software: the {\tt \lowercase{pylspack}} package\label{sec:software}}
An implementation of the proposed algorithms, whose core is written in C++ and parallelized with OpenMP \cite{dagum1998openmp} for multithreading and SIMD directives, is publicly available on Github. {Python bindings are also provided, compatible with standard numerical data structures of NumPy and SciPy}. We list the basic endpoints of the Python API with a short description. Hereafter, ${\tt csr\_matrix}$ refers to ${\tt scipy.sparse.csr\_matrix}$ data structure, ${\tt ndarray}$ refers to ${\tt numpy.ndarray}$, ${\tt C\_CONTIGUOUS}$ is a ${\tt ndarray}$ stored in row-major ordering and ${\tt F\_CONTIGUOUS}$ is in column-major ordering. The size of the input matrix $A$ in all of the following descriptions is $n\times d$.

\begin{description}
\item[${\tt csrcgs}(A, m, r)$:]
Given a ${\tt csr\_matrix}$ $A$ and integers $m$ and $r$, compute the product $GSA$ where G is a $m\times r$ matrix with standard normal random elements scaled by $1/\sqrt{m}$ and S is a $r\times n$ CountSketch transform (by setting $m=0$, only SA is computed and returned). The result is ${\tt C\_CONTIGUOUS}$. This is an implementation of Algorithm \ref{alg:countgauss_multiply}.
\item[${\tt csrjlt}(A, m)$:] Given a ${\tt csr\_matrix}$ A and an integer $m$, compute the product GA where G is as above. The result is ${\tt F\_CONTIGUOUS}$ ${\tt ndarray}$. This is an implementation of Algorithm \ref{alg:gaussian_sketch}.
\item[${\tt csrrk}(\alpha, A, \beta, C)$:] Given scalars $\alpha$ and $\beta$, a ${\tt csr\_matrix}$ A and a ${\tt C\_CONTIGUOUS}$ ${\tt ndarray}$ $C$, computes $C\leftarrow \alpha A^\top A + \beta C$ ($C$ is updated inplace). This is an implementation of Algorithm \ref{alg:gram}.
\item[${\tt csrsqn}(\alpha, A, B, \beta, x)$:] Given scalars $\alpha$ and $\beta$, a ${\tt csr\_matrix}$ $A$ and a ${\tt C\_CONTIGUOUS}$ ${\tt ndarray}$ $B$, computes $x \leftarrow \alpha y + \beta x$, where $y$ is a vector of size $n$ containing the squared Euclidean norms of the rows of the matrix $C = A B$. Neither $C$ nor $y$ are formed explicitly. This is an implementation of Algorithm \ref{alg:squared_row_norms}.
\item[${\tt rmcgs}(A, m, r)$:] Same as ${\tt csrcgs}$, but $A$ a ${\tt C\_CONTIGUOUS}$ ${\tt ndarray}$ in this case. This method implements a variant of Algorithm \ref{alg:countgauss_multiply} tuned for a ${\tt C\_CONTIGUOUS}$ input matrix.
\item[${\tt rmsqn}(\alpha, A, B, \beta, x)$:] Same as ${\tt csrsqn}$, but in this case A is a ${\tt C\_CONTIGUOUS}$ ${\tt ndarray}$. This is a memory efficient implementation of Algorithm \ref{alg:squared_row_norms} tuned for dense $A$.
\end{description}

The OpenMP behaviour is fully controllable via the standard environment variables such as ${\tt OMP\_NUM\_THREADS}$ and ${\tt OMP\_PLACES}$. Based on the aforementioned functions as well as the ones provided by SciPy and NumPy, the package also contains implementations for column subset selection and leverage scores as described the previous sections. {Function calls} for least squares solvers and preconditioning are not explicitly available, but they can be easily derived using just a few calls of the existing ones. For the following functions, $A$ can be either a ${\tt csr\_matrix}$ or a ${\tt C\_CONTIGUOUS}$ ${\tt ndarray}$.

\begin{description}
\item[${\tt sample\_columns}(A, rcond, m, r)$:] Select a subset of columns of $A$ based on Algorithm \ref{alg:cgrrqr}. Here $m$ and $r$ define the parameters for the underlying call to ${\tt csrcgs/rmcgs}$, and rcond is used as a threshold for the small singular values to determine the numerical rank k and the subspace dimension.
\item[${\tt ls\_via\_inv\_gram}(A, rcond)$:] Compute the leverage scores of the best rank-k approximation of A, which is determined based on rcond. This is an implementation of Algorithm \ref{alg:ls_direct}.
\item[${\tt ls\_via\_sketched\_svd}(A, rcond, m, r1, r2)$:] Approximately compute the leverage scores of the best rank-k approximation of A using ${\tt rmcgs/csrcgs}$. Here $m, r1, r2$ define the sketching dimensions. This is an implementation of Algorithm \ref{alg:ls_approximate_drineas}.
\item[${\tt ls\_hrn\_exact}(A, rcond, m, r)$:] Approximate the leverage scores of A using ${\tt sample\_columns}$ as a first step and then calling ${\tt ls\_via\_inv\_gram}$ on the selected column subset. This is an implementation of Algorithm \ref{alg:ls_hrn_exact}.
\item[${\tt ls\_hrn\_approx}(A, rcond, m, r, m\_ls, r1\_ls, r2\_ls)$:] Approximate the leverage scores of A using ${\tt sample\_columns}$ as a first step and then calling ${\tt ls\_via\_sketched\_svd}$ on the selected column subset. This is an implementation of Algorithm \ref{alg:ls_hrn_exact}.
\end{description}

In Table \ref{tab:basic_kernels_requirements} we list the number of flops that are required per operation as well as the memory requirements for temporary storage (``aux memory'' column) and the total of random numbers that need to be generated during execution (``rng'' column).

\begin{table}
\scriptsize
    \centering
    \caption{Implementation characteristics for the basic algorithms of the {\tt pylspack} package. Here $A\in\mathbb{R}^{n\times d}$, $m$ and $r$ are input parameters and $p$ is the number of OpenMP threads. $|{\tt int}|$ and $|{\tt double}|$ denote the number of words required to store integers and double precision floating point numbers respectively.}
    \begin{tabular}{l | c | c | c}
        operation & flops & rng & aux memory \\\hline
        {\tt csrcgs} & {${\tt nnz}(A)+n+(2r-1)md$} & $mr\times {\tt randn}$ + $n\times {\tt randi}$ + $n\times {\tt randb}$ & $pd\times |{\tt int}| + (m+d)d \times |{\tt double}|$\\
        {\tt csrjlt} & $2{\tt nnz(A)}m $ & $mn\times {\tt randn}$ &$m\times |{\tt double}|$\\
        {\tt csrrk} & $2{\tt nnz_2}(A)+{\tt nnz}(A)+d^2$ & - & - \\
        {\tt csrrk} & $2{\tt nnz_2}(A)+{\tt nnz}(A)+d^2$ & - & - \\
        {\tt csrsqn} & $3{\tt nnz_2}(A)+{\tt nnz}(A)+2n+(2r-1)d^2$ & - & - \\
        {\tt rmcgs} & {$nd+n+(2r-1)md$} & $mr\times {\tt randn}$ + $n\times {\tt randi}$ + $n\times {\tt randb}$ & $pd\times |{\tt int}| + (m+d)d \times |{\tt double}|$ \\
        {\tt rmsqn} & $2ndr-nr$  & - & - \\
    \end{tabular}
    \label{tab:basic_kernels_requirements}
\end{table}

\subsection{Pseudo-random number generation}
For pseudo-random number generation, we used existing methods from {STL} {and in particular:}
\begin{itemize}
    \item {\tt  default\_random\_engine} + {\tt bernoulli\_distribution} for {\tt randb},
    \item {\tt mt19937} + {\tt uniform\_int\_distribution} for {\tt randi},
    \item {\tt mt19937\_64} + {\tt normal\_distribution} for {\tt randn}. 
\end{itemize}
{Each thread constructs its own random engine object, instantiated with a different seed based on the thread id. The current version of the code uses this simple solution which also avoids any external dependencies. In future work we plan to consider specialized parallel random number generators, like the Random123 library \cite{salmon2011parallel} which is used by Skylark.}

\subsection{Unit Testing}
{
{\tt pylspack} is accompanied with an extensive test suite. Each kernel is tested for a large variety of possible input parameters, input sizes and sparsity patterns, both random and deterministic. All kernels are evaluated in terms of the (forward) floating point errors, while the random projection kernels are also tested with respect to their ability to satisfy certain criteria, e.g. $\epsilon$-OSE/JL. For the advanced algorithms, namely for leverage scores computation, column subset selection, rank computation, there are also more advanced tests, i.e., it is tested that the rank approximation method is actually returning the correct rank. Respectively, the leverage scores computation algorithms are tested in terms of their ability to satisfy the probabilistic approximation guarantees, and the column subset selection algorithms in terms of their ``rank-revealing'' properties. 
In the automated pipelines, we test all configurations possible from the combination of the following: two different compilers, three different Python versions, single threaded and multi-threaded execution. We used the ${\tt pytest}$ Python package in order to facilitate our tests.
}

\section{Experiments\label{sec:experiments}}

We next compare the performance of the {\tt pylspack} with existing {state-of-the-art} numerical libraries. Specifically, we compare the sketching methods with the corresponding ones provided in Scikit-learn and Skylark. Scikit-learn was installed via Pypi, while Skylark was installed with the provided installer scripts from the main website.\footnote{{https://xdata-skylark.github.io/libskylark/}{https://xdata-skylark.github.io/libskylark/}}
{
For the computation of the Gram matrix we compare with the SciPy built-in functions as well as with the {\tt mkl\_sparse\_?\_syrkd} method of Intel MKL. Note, however, that as of the time of writing there exist no official Python bindings for Intel MKL. For our experiments we used the ones available publicly through the ${\tt sparse\_dot\_mkl}$ Python package \cite{sparse_dot_mkl}.} 
{Hereafter, we will refer to the Python package of Scikit-learn as ${\tt sklearn}$, the Python package of Skylark as ${\tt skylark}$, and the Python package of SciPy as ${\tt scipy}$. In general, we will use normal font to refer to the project names, and typescript for the respective Python packages. 
We need to note that ${\tt pylspack}$ is parallel by design and exploits multithreading. This is not the case with most components of the packages we are comparing with. This is an advantage of ${\tt pylspack}$. In the interest of fairness, this will be taken into account in the experimental evaluations.
}
We installed {\tt pylspack}, ${\tt sklearn}$, ${\tt scipy}$ and ${\tt sparse\_dot\_mkl}$ inside the same Python 3  environment, such that they share the same dependencies, e.g. SciPy, NumPy, OpenBLAS etc. Skylark is implemented in Python 2 and it is shipped with its own Python 2 environment, including all the dependencies. Therefore, unavoidably, the ${\tt skylark}$ package is linked with different dependencies than the other libraries. {MKL, which is used as the backend of {sparse\_dot\_mkl}, was installed via the official Anaconda repositories of Intel.}

{We perform four different types of experiments, by grouping together methods that perform similar tasks. In all cases we compare the performance for sparse input matrices, and, where applicable, we also compare for dense input matrices. The experiments were conducted on a machine with $128$ GB of RAM and one single-socket AMD EPYC processor, with $16$ physical cores and supporting up to $2$ hyperthreads per core, that is a total of $32$ parallel threads. 
}

\subsection{Test matrices and parameters}
{For all experimental setups, we test two different matrix sizes (same size both for sparse and dense). The first matrix size is $2,097,152\times 512$, and it sparse version has $5\%$ density, that is, $95\%$ of its elements are zero. In terms of dimensions, this latter sparse matrix has similar characteristics to the ``TinyImages'' matrix \cite{sobczyk2021estimating,meng2014lsrn}, originating from the TinyImages dataset \cite{torralba200880}.  Although ${\tt pylspack}$ primarily targets tall-and-skinny matrices, to better appreciate its performance we also conduct experiments with  ``shorter-and-wider'' matrices of size $131,072\times 8,192$, one dense and the other sparse with density $ 2.5\%$. All the matrices used in the experiments are listed in Table \ref{tab:matrices}.}

{For the sketching methods, we define a parameter $k\leq d$, which is the dimension of the target subspace to be approximated. This is in order to evaluate performance when one  is interested in approximating only the leading $k$-dimensional subspace of the input matrix, rather than the full column subspace; cf. \cite{woodruff2014sketching,meyer2021hutch++}. Moreover, the shorter-and-wider matrix does not fulfill the size requirements for a CountSketch subspace embedding, since $(8,192)^2\gg 131,072$.\footnote{Even by relaxing the subspace embedding requirement of CountSketch from $d^2$ to $100d$, for the $131,072\times 8,192$ sized matrix we have that $100d=100\times 8,192\gg 131,072$.} For the tall matrix, we choose $k=d=512$, while for the shorter-and-wider matrix we choose $k=100$. We used the {\tt pytest}-{\tt benchmark} Python package to facilitate our experiments. Each method is run $15$ times, split in $3$ rounds of $5$ iterations per round. In Table \ref{tab:method_acronyms} we summarize the names/acronyms that we will use to refer to the different methods considered.}
\begin{table}[htb]
    \caption{Matrices used in exoperiments. $n$ is the number of rows and $d$ is the number of columns.}
    \centering
    \begin{tabular}{r r r c}\hline
    Description & $n$ & $d$ & density\\\hline\hline
    Tall dense  & $2,097,152$ & $512$ & fully dense \\
    Short dense  & $131,072$ & $8,192$ & fully dense \\
    Tall sparse  & $2,097,152$ & $512$ & $5\%$ nonzeros \\
    Short sparse  & $131,072$ & $8,192$ & $2.5\%$ nonzeros \\\hline
    \end{tabular}
    \label{tab:matrices}
\end{table}

\begin{table}[htb]
    \centering
    \footnotesize
    \caption{{Names and acronyms used for different methods in the experiments. For multithreading, ${\tt skylark}$ requires sparse storage in the CSC format.}}
    {
    \begin{tabular}{l | l | l}
    \hline
    name/acronym & actual function & multithreaded \\\hline\hline
        ${\tt csrrk}$ &  ${\tt pylspack.linalg\_kernels.csrrk}$ & yes (CSR)\\
        ${\tt csrjlt}$ &  ${\tt pylspack.linalg\_kernels.csrjlt}$ & yes (CSR)\\
        ${\tt csrcgs}$ &  ${\tt pylspack.linalg\_kernels.csrcgs}$ & yes (CSR)\\
        ${\tt csrsqn}$ &  ${\tt pylspack.linalg\_kernels.csrsqn}$ & yes (CSR) \\\hline
        ${\tt JLT}$ &  ${\tt skylark.sketching.JLT}$ & yes (CSC)\\
        ${\tt CWT}$ &  ${\tt skylark.sketching.CWT}$ & yes (CSC)\\\hline
        ${\tt GRP}$ &  ${\tt sklearn.random\_projection.GaussianRandomProjection}$ & no\\
        ${\tt SRP}$ &  ${\tt sklearn.random\_projection.SparseRandomProjection}$ & no\\\hline
        ${\tt GM}$-${\tt MKL}$ &  ${\tt sparse\_dot\_mkl.gram\_matrix\_mkl}$, Python wrapper for {\tt mkl\_sparse\_?\_syrkd} & yes\\\hline
    \end{tabular}    }
    \label{tab:method_acronyms}
\end{table}

\subsection{Gaussian sketching}
{The first type of experiment concerns Gaussian random projections. The following methods are compared: ${\tt csrjlt}$, ${\tt GRP}$ and ${\tt JLT}$. We run the experiments for sparse input matrices which are stored in the CSR format in the case of ${\tt pylspack}$ and ${\tt sklearn}$, and in the CSC format in ${\tt skylark}$. In all cases the output is stored as a dense matrix. In Figure \ref{fig:jlt_tall} we plot the runtimes for the aforementioned methods over $1,2,4,8,16$ and $32$ threads. Up to $16$ threads there is one thread per physical core. For $32$ threads, two hyperthreads per core are launched.  For both matrices, the number of rows $m$ of $G$ is set to be equal to $2k$, where $k$ is the dimension of the target subspace to be approximated. For the tall matrix, we set $k=d=512$, while for the short matrix the target dimension is set to $k=100$. Method ${\tt csrjlt}$ is clearly faster than both of the other methods and scales better as the number of threads increases. We also observe that, for the tall matrix, the ${\tt JLT}$ method is slightly faster on a single thread, but its performance recedes as the number of threads increases. For the short matrix, ${\tt csrjlt}$ shows a performance degradation at $p=4$ threads. This is attributed to the processor architecture and the sparse, memory-bound nature of the sparse matrix multiplication. Such phenomena can occur once the threads start sharing higher cache levels, especially for such short total runtimes. To address such issues, however, is beyond the scope of this paper.
\begin{figure}[htb]
    \centering
    \includegraphics[width=0.45\textwidth]{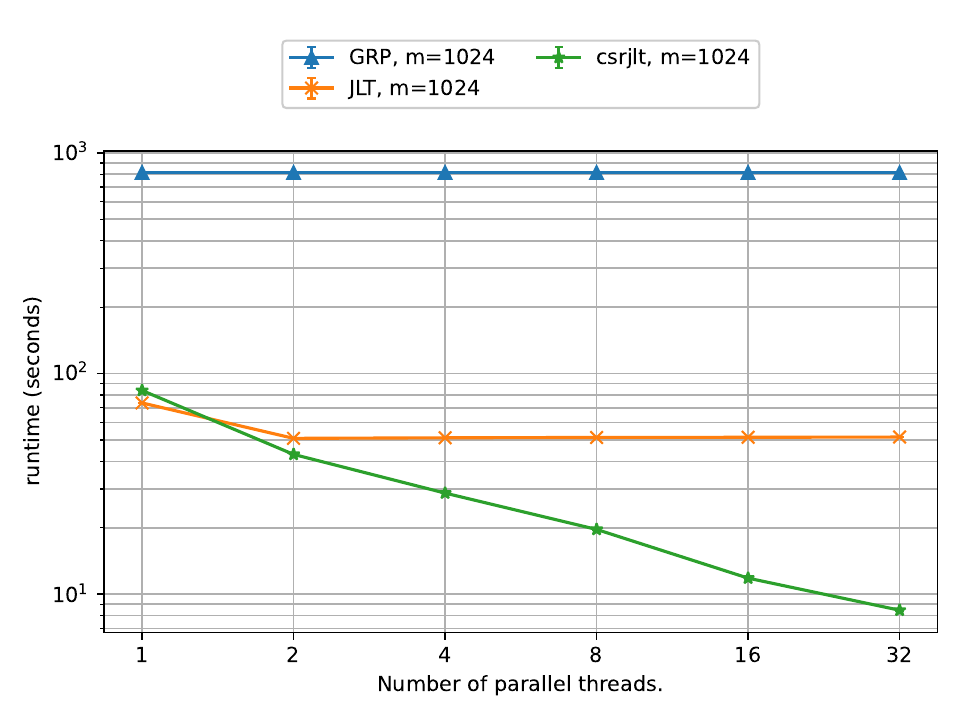}
    \includegraphics[width=0.45\textwidth]{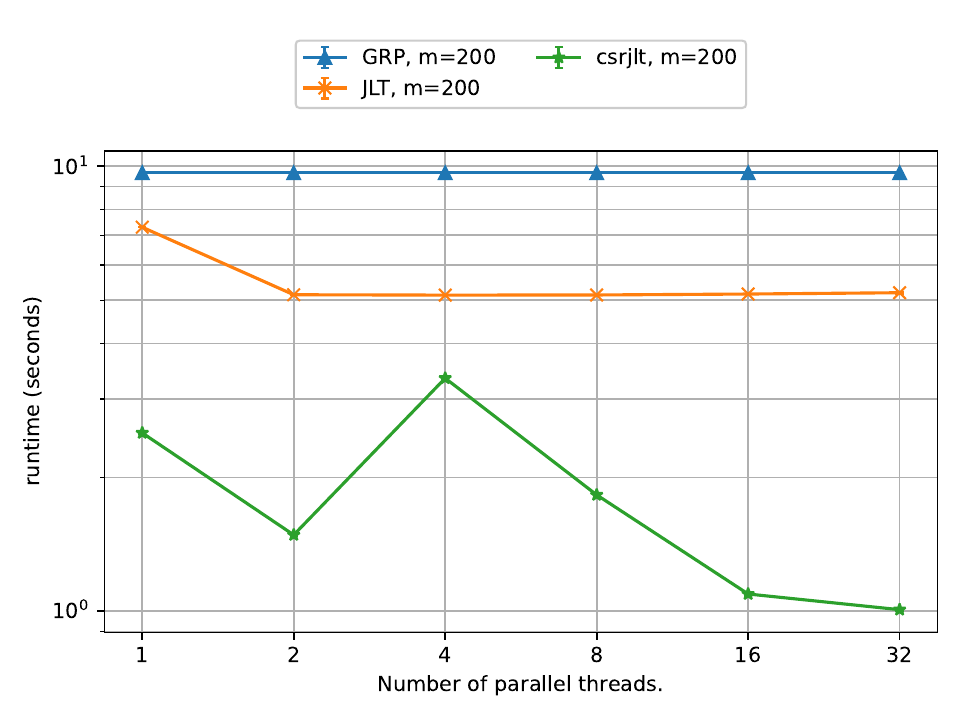}
    \caption{Runtime comparison for Gaussian random projections. Each method computes the product $GA$, where $A$ is the input matrix and $G$ is a Gaussian matrix. $m$ denotes the number of rows of $G$. {\bf Left}: Tall sparse matrix with size $2M\times 512$ and density $\sim 5\%$. {\bf Right}: Sparse rectangular matrix with size $131K\times 8K$ and density $\sim 2.5\%$.}
    \label{fig:jlt_tall}
\end{figure}
}

\subsection{Sparse sketching}
{We next compare the corresponding routines of the ${\tt sklearn}$, ${\tt skylark}$ and ${\tt pylspack}$ Python packages for sparse sketching, namely ${\tt csrcgs}$, ${\tt CWT}$ and ${\tt SRP}$. All methods perform the computation of the product $SA$ where $S$ is a sparse random projection matrix with $r$ rows, and $A$ is the input matrix. The first two implement a CountSketch transform, while the third method implements a a sparse Rademacher type of transform for $S$. For this type of experiment we consider both sparse and dense matrices. For dense matrices, ${\tt rmcgs}$ is used instead of ${\tt csrcgs}$. The number of rows $r$ of $S$ is set to be equal to $10k$, where $k$ is the target dimension that needs to be approximated. For the tall matrix we set $k=d=512$, while for the short matrix we set $k=100$. }

{For both the sparse and the dense input matrices, and for all matrix sizes, the methods of ${\tt pylspack}$ are faster than the other methods and scale better over the number of threads. In particular, for the maximum $32$ threads, it is faster by more than one order of magnitude. We recall, however, that the methods of ${\tt sklearn}$ are not multithreaded, and that ${\tt skylark}$  targets distributed computations rather than shared memory parallelism.}

{
\begin{figure}[H]
    \centering
    \includegraphics[width=0.45\textwidth]{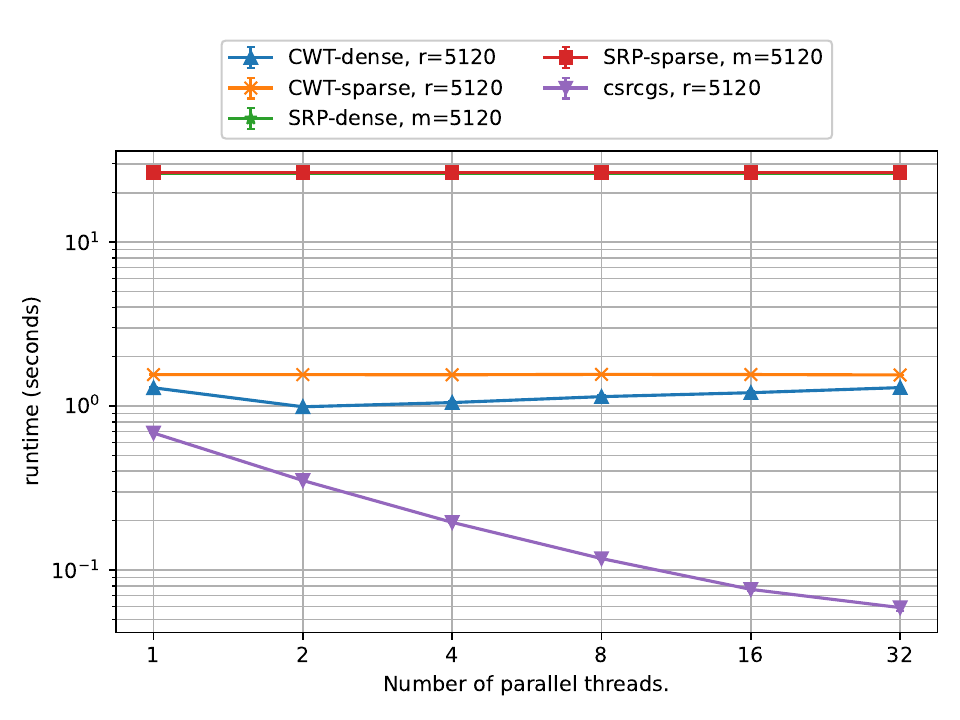}
    \includegraphics[width=0.45\textwidth]{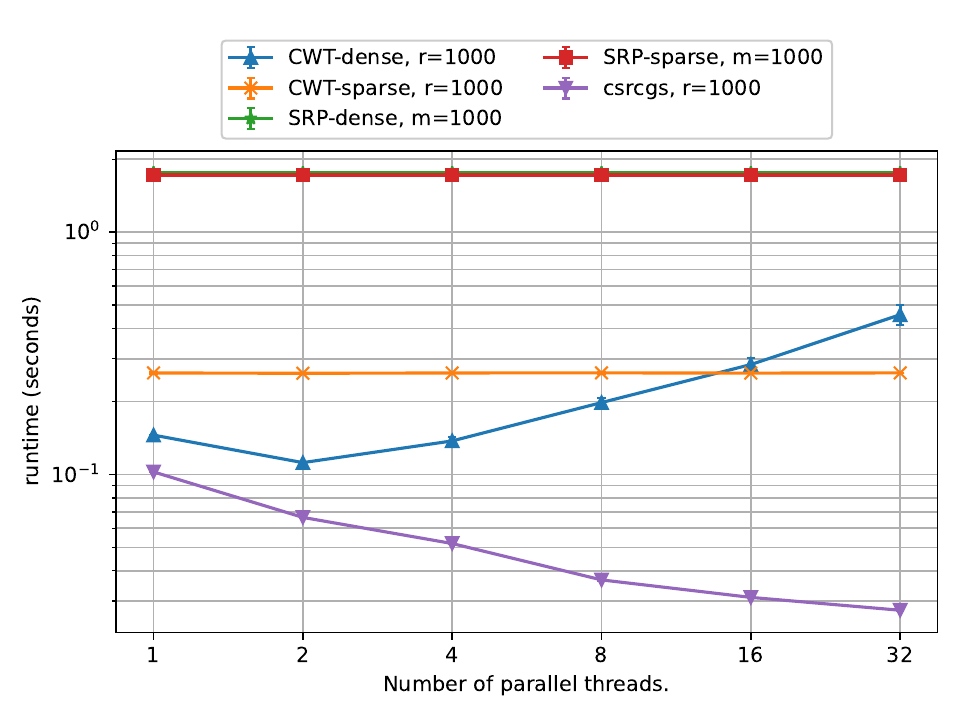}
    \caption{Runtime comparison for sparse random projections. Each method computes the product $SA$, where $A$ is a sparse input matrix and $S$ is a CountSketch matrix for ${\tt csrcgs}$ and for ${\tt CWT}$, while for ${\tt SRP}$ it is a sparse Rademacher transform. $r$ denotes the number of rows of $S$. The ``-${\tt sparse}$'' and ``-${\tt dense}$'' suffixes in the legend denote that the output is stored in dense/sparse format, respectively. {\bf Left}: Tall sparse matrix with size $2M\times 512$ and density $\sim 5\%$. {\bf Right}: Sparse rectangular matrix with size $131K\times 8K$ and density $\sim 2.5\%$.}
    \label{fig:sparse_tall_sparse}
\end{figure}
\begin{figure}[htb]
    \centering
    \includegraphics[width=0.45\textwidth]{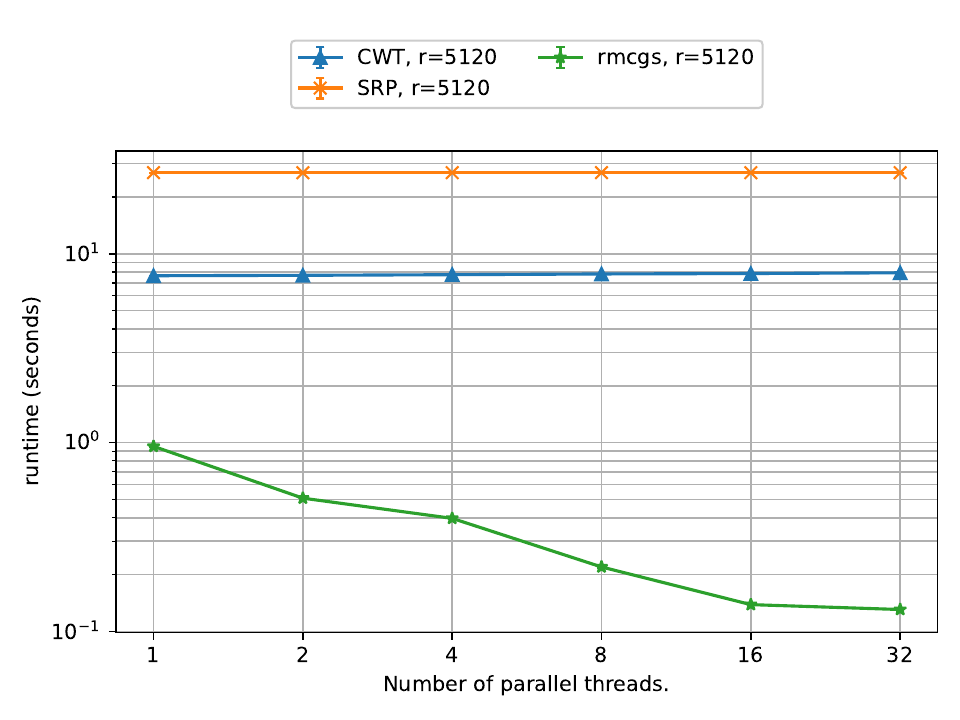}
    \includegraphics[width=0.45\textwidth]{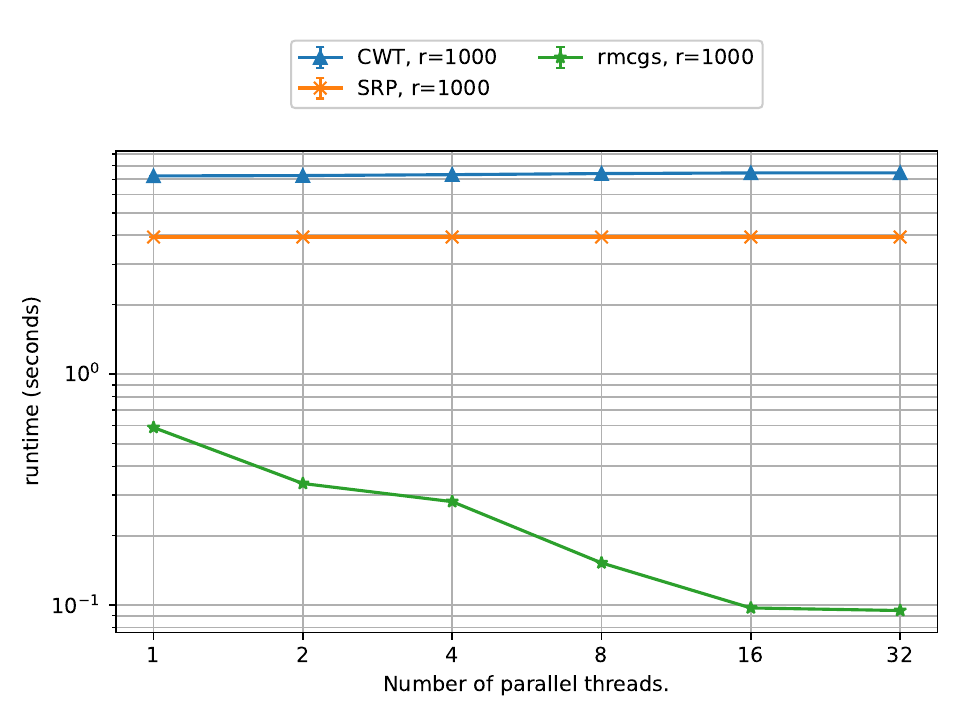}
    \caption{Runtime comparison for sparse random projections. Each method computes the product $SA$, where $A$ is a dense input matrix and $S$ is a CountSketch matrix for ${\tt csrcgs}$ and for ${\tt CWT}$, while for ${\tt SRP}$ it is a sparse Rademacher transform. $r$ denotes the number of rows of $S$. {\bf Left}: Tall dense matrix with size $2M\times 512$. {\bf Right}: Dense rectangular matrix with size $131K\times 8K$.}
    \label{fig:sparse_tall_dense}
\end{figure}
}

\subsection{Multi-level sketching}
{We next consider the computation of the product $GSA$ where $A$ is the input matrix, $S$ is a sparse projection as above, and $G$ is a Gaussian random projection. Both sparse and dense matrices are considered for these experiments. We compare the method ${\tt csrcgs}/{\tt rmcgs}$ from ${\tt pylspack}$ for sparse/dense inputs accordingly. For ${\tt sklearn}$, we first apply ${\tt SRP}$ on the input and then ${\tt GRP}$ on the result. Similarly, for ${\tt skylark}$ we first apply ${\tt CWT}$ on the input matrix and then ${\tt JLT}$ on the result. $r$ denotes the number of rows of $S$ as well as the columns of $G$, and $m$ denotes the number of rows of $G$. As in the previous experiments, we define $k$ to be the dimension of the target subspace. For the tall matrix we set $k=d=512$, and for the short matrix we set $k=100$. We then set $r=100k$ for the rows of $S$ and $m=2k$ for the rows of $G$. ${\tt csrcgs}$ and ${\tt rmcgs}$ are many times faster and scale a lot better over the number of threads as compared to the other methods. Moreover, they are much more memory efficient because of their batched organization, which requires only $O(pd+md)$ storage for intermediate results, instead of $O(rd)$ that is required by the other two methods. For the tall matrix, $SA$ requires roughly $51200\times 512\times 8 / 10^6\approx200$ MB, while ${\tt csrcgs}/{\tt rmcgs}$ require about $1024\times 512\times 8/10^6\approx 4.2$ MB. For the short matrix, at $32$ threads, hyperthreading takes place, the threads need to share resources, and the performance of ${\tt csrcgs/rmcgs}$ deteriorates. We attribute this to the dense matrix-multiplication operation used by the aforementioned methods, which we opted to implement ourselves instead of relying on external libraries introducing unnecessary dependencies, which we felt were not justified. Indeed, note, that the runtime in this case is only about $300$ ms.
\begin{figure}[H]
    \centering
    \includegraphics[width=0.45\textwidth]{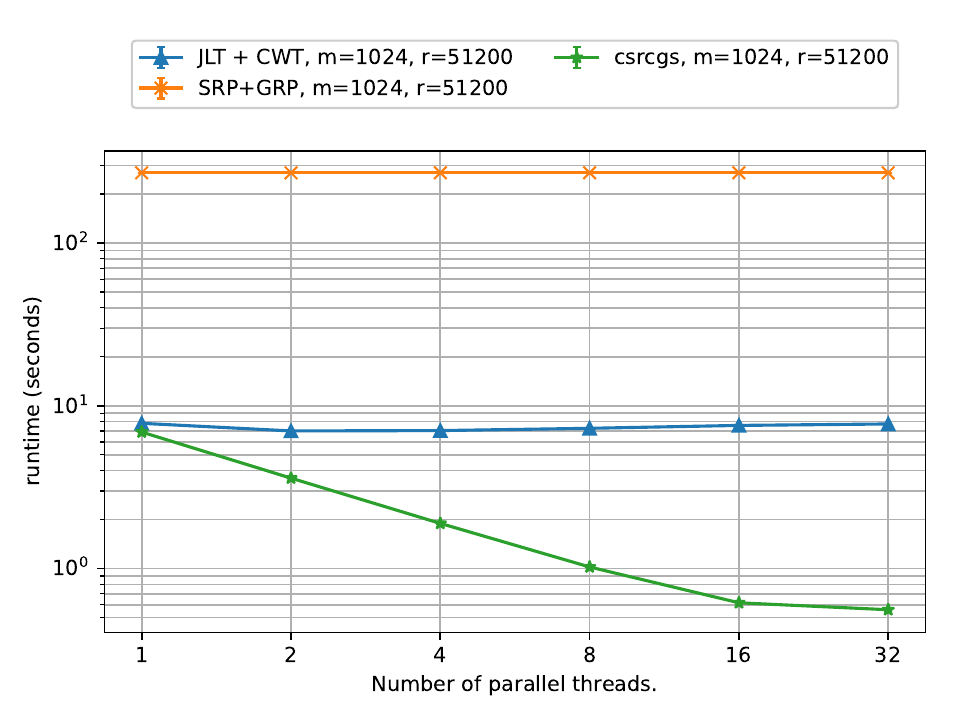}
    \includegraphics[width=0.45\textwidth]{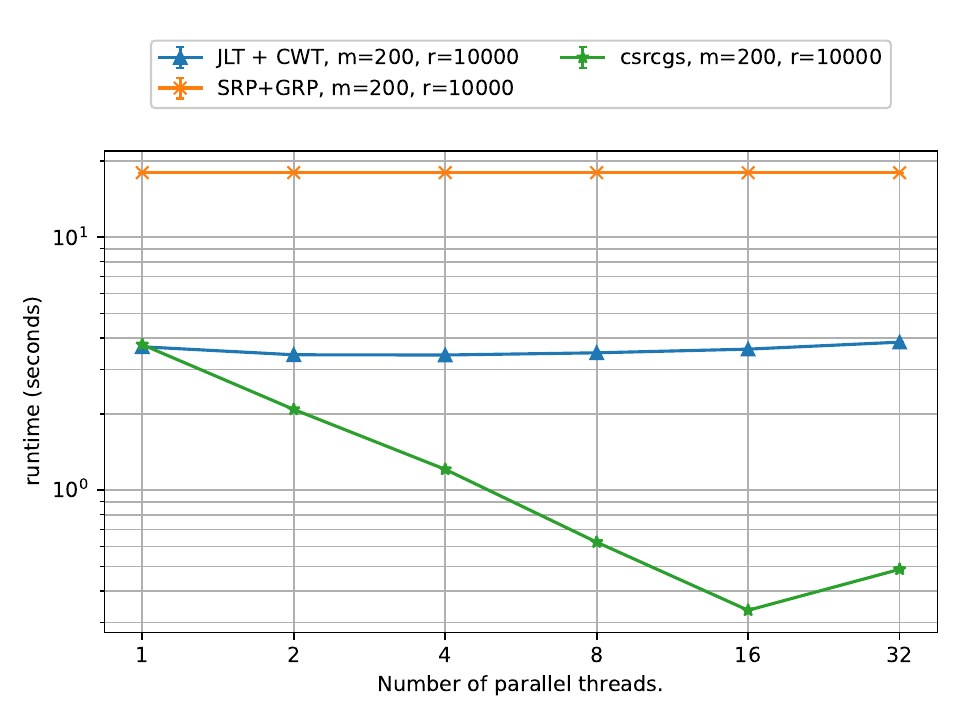}
    \caption{Runtime comparison for combined sparse and Gaussian random projections. Each method computes the product $GSA$, where $A$ is a sparse input matrix, $S$ is a CountSketch matrix for ${\tt csrcgs}$ and for ${\tt CWT}$, while for ${\tt SRP}$ it is a sparse Rademacher transform, and $G$ is a Gaussian matrix. $r$ denotes the number of rows of $S$ (and the number of columns of $G$), and $m$ denotes the number of rows of $G$. All the resulting matrices are stored in dense format. {\bf Left}: Tall sparse matrix with size $2M\times 512$ and density $\sim 5\%$. {\bf Right}: Sparse rectangular matrix with size $131K\times 8K$ and density $\sim 2.5\%$.}
    \label{fig:combined_tall_sparse}
\end{figure}
\begin{figure}[H]
    \centering
    \includegraphics[width=0.45\textwidth]{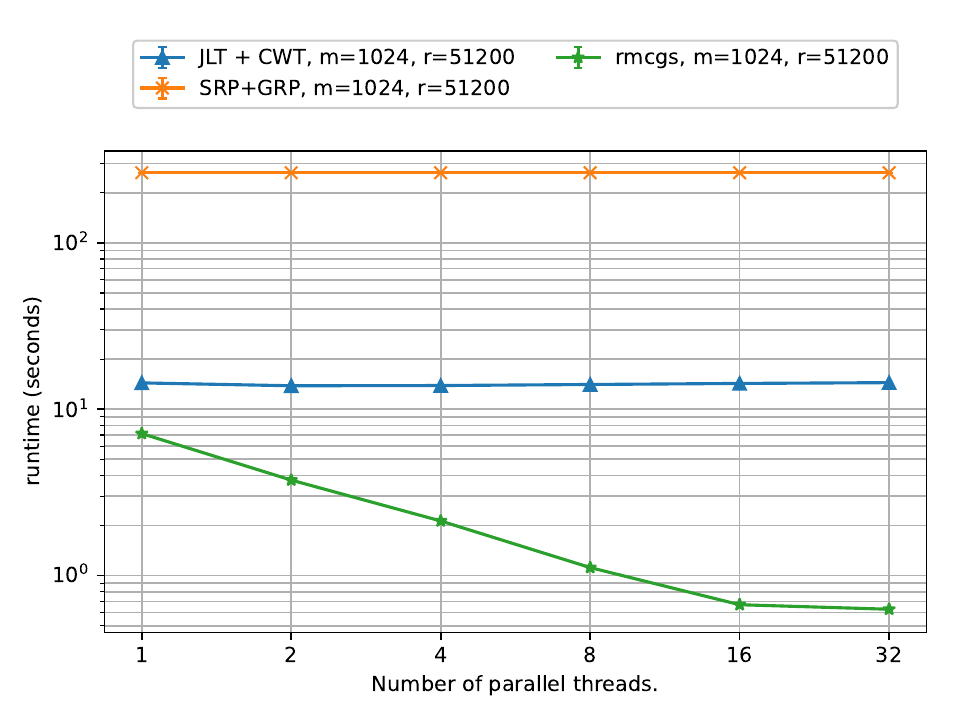}
    \includegraphics[width=0.45\textwidth]{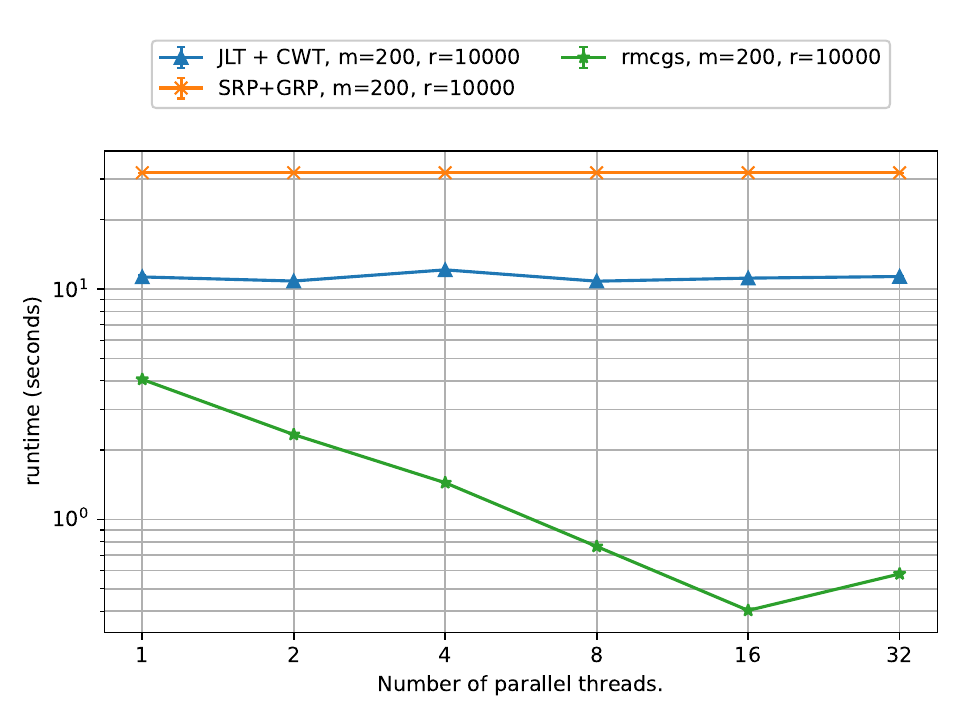}
    \caption{Runtime comparison for combined sparse and Gaussian random projections. Each method computes the product $GSA$, where $A$ is a dense input matrix, $S$ is a CountSketch matrix for ${\tt csrcgs}$ and for ${\tt CWT}$, while for ${\tt SRP}$ it is a sparse Rademacher transform, and $G$ is a Gaussian matrix. $r$ denotes the number of rows of $S$ (and the number of columns of $G$), and $m$ denotes the number of rows of $G$. All the resulting matrices are stored in dense format. {\bf Left}: Tall dense matrix with size $2M\times 512$. {\bf Right}: Dense rectangular matrix with size $131K\times 8K$.}
    \label{fig:combined_tall_dense}
\end{figure}
}

\subsection{Gram matrix computation}
{We next compare SciPy, sparse\_dot\_mkl and ${\tt pylspack}$ for the computation of the Gramian $A^\top A$ when $A$ is sparse. Results are shown in Figure \ref{fig:sgram_tall_sparse}. We also opt to plot the performance of ${\tt csrsqn}$ in the same figure. The plots were combined in the interest of economy of the presentation because even though ${\tt csrsqn}$ corresponds to a different operation, the runtimes are of the same magnitude. SciPy does not offer the possibility to store the resulting matrix as a dense matrix. ${\tt sparse\_dot\_mkl}$ offers both sparse and dense output. We will refer as ${\tt GM}$-${\tt MKL}$-${\tt sparse}$ to the ${\tt GM}$-${\tt MKL}$ method for the sparse output case, and as ${\tt GM}$-${\tt MKL}$-${\tt dense}$  for the same method for dense output.
}

{
The performance plots in this case are quite intriguing. For both the tall and the short matrix, the ${\tt GM}$-${\tt MKL}$-${\tt dense}$ method displays a performance degradation starting from two threads. This indicates that the single-threaded and multi-threaded implementation of that specific function are substantially different. For the tall matrix, the single threaded execution is faster than the multi-threaded, even for $16$ and $32$ threads. On the other hand, the ${\tt GM}$-${\tt MKL}$-${\tt sparse}$ method behaves much better. Interestingly, the built-in matrix multiplication method of ${\tt scipy}$ is only slightly slower than this method for the single-threaded case, for both matrices. For the short matrix, ${\tt scipy}$ is interestingly slightly faster than ${\tt GM}$-${\tt MKL}$-${\tt dense}$ and  than ${\tt csrrk}$. For the tall-and-skinny case, ${\tt csrrk}$ is by far the fastest method, being over one order of magnitude faster than the others for $32$ threads, even the highly optimized ones of Intel MKL. This arguably proves that a dedicated implementation for the computation of the Gramian for tall-and-skinny matrices is well justified.  ${\tt csrrk}$ also exhibits good performance  for the short matrix, though in this case, ${\tt GM}$-${\tt MKL}$-${\tt sparse}$ method is about twice as fast, while the scaling behaviour of both methods is similar. It is worth noting, however, that the output matrix is dense, thus storing it using a sparse format, as in ${\tt GM}$-${\tt MKL}$-${\tt sparse}$, entails significant redundancy. We conclude that that ${\tt csrrk}$ should be the method of choice when the resulting Gramian is expected to be dense.
}

{
\begin{figure}[htb]
    \centering
    \includegraphics[width=0.45\textwidth]{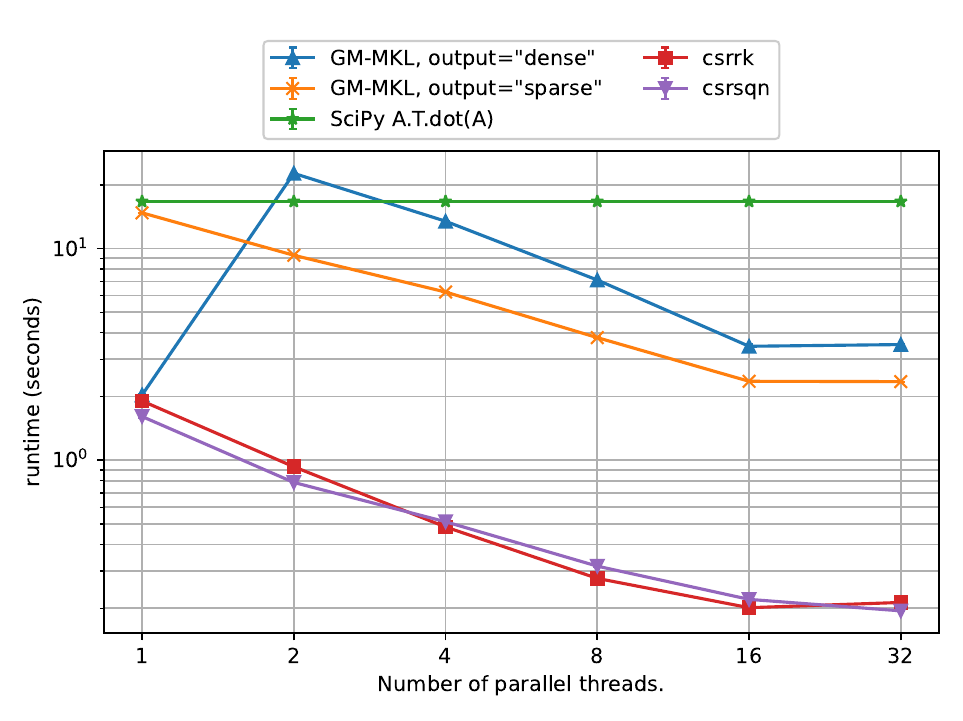}
    \includegraphics[width=0.45\textwidth]{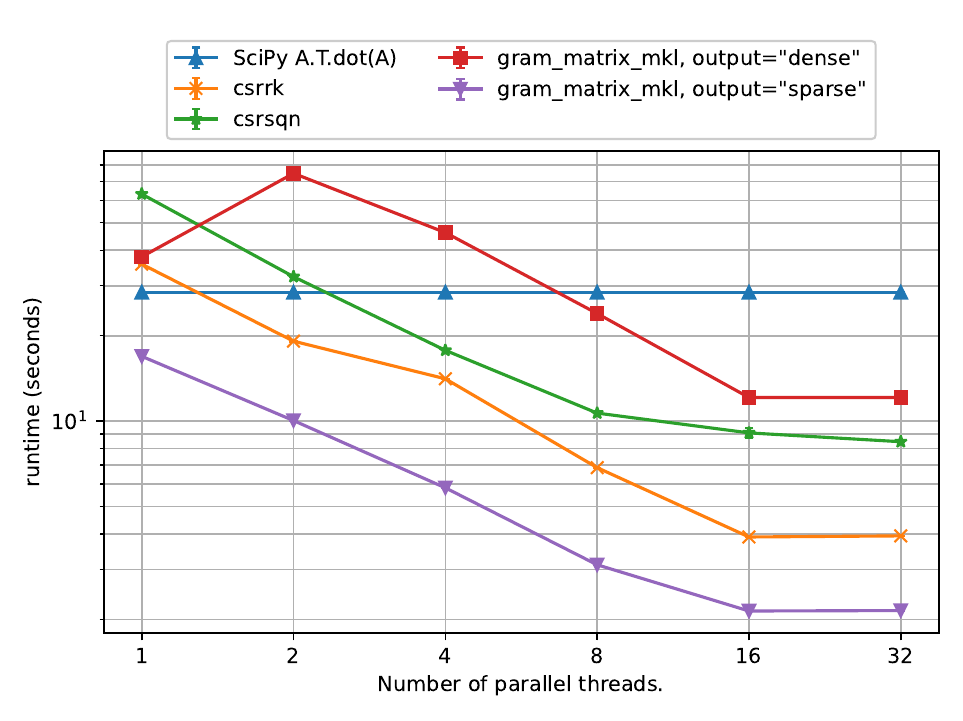}
    \caption{Runtime comparison for the computation of the Gramian $A^\top A$ of a sparse input matrix $A$ (${\tt csrsqn}$ performs a different operation, but it is added in this plot as it has similar complexity). The built-in multiplication of SciPy returns the result in sparse format. {\bf Left}: Tall sparse matrix with size $2M\times 512$ and density $\sim 5\%$. {\bf Right}: Sparse rectangular matrix with size $131K\times 8K$ and density $\sim 2.5\%$.}
    \label{fig:sgram_tall_sparse}
\end{figure}
}

\section{Conclusions\label{sec:conclusions}} 

We described parallel algorithms and data structures for {Gaussian} and  {CountSketch random projection methods} {as well as the computation of the Gram matrix} and their application to column subset selection, leverage scores and least squares regression. We have shown that {the proposed} methods can be implemented efficiently and provided detailed description of our open source {\tt pylspack} Python package. We showed that the algorithms scale well on a shared memory machine and that they can outperform other existing open-source and vendor libraries for the same tasks in terms of both performance and memory requirements.

{For the sketching operation with a CountSketch matrix, we showed that large fluctuations in the expected workload distribution among processors occur with only small probability. These results can provide the basis for a more detailed analysis of load balancing and the design of more aggressive algorithms addressing this issue.}
Other topics that could be considered for future improvements are specialized compression schemes \cite{kourtis2008optimizing,kourtis2011csx} for the sparse matrices, as well as the choice or random number generators. For standard normal variables, for example, the Ziggurat method \cite{marsaglia2000ziggurat} can be substantially faster than the STL random number generators; cf. \cite{meng2014lsrn}.

{We conclude with the following observation. As already discussed, to compute the leverage scores, one can compute the Gramian $A^\top A$ and its pseudoinverse $B$, and then compute the row norms of $AB^{\frac{1}{2}}$, all in $O({\tt nnz_2}(A)+d^\omega)$. However, one cannot do the ``opposite'', i.e. there is no reduction from the Gramian computation to the computation of leverage scores, or any other reduction in general between the aforementioned problems. In light of recent advances in the Fine-Grained complexity literature \cite{kyng2020packing,bafna2021optimal,musco2018spectrum}, finding such reductions would be an interesting and challenging problem. }

\section*{Acknowledgements}
We thank George Kollias for advising us regarding installation and implementation details of Skylark, Zoltan Arnold Nagy for letting us access the necessary computing resources, Yousef Saad for comments on SPARSKIT and Gram matrix computation and Vasilis Georgiou for discussions on SpGEMM and the Ginkgo library. We also thank anonymous reviewers for their comments and criticism which helped improve a previous version of this manuscript.

\bibliographystyle{plain}

\end{document}